\documentclass[a4paper,fleqn]{cas-dc}
\usepackage[numbers]{natbib} 
\setcitestyle{numbers} 
\usepackage{graphicx}
\usepackage{array}
\usepackage{setspace}
\usepackage[T1]{fontenc}
\usepackage[utf8]{inputenc}   
\usepackage{textcomp}
\usepackage{microtype}
\usepackage{amsmath,amssymb,bm,xfrac,mathtools,amsthm}
\usepackage[export]{adjustbox}
\usepackage{hhline}
\usepackage{tikz} \usetikzlibrary{calc,shapes,arrows,automata,positioning} 
\usepackage{acronym}
\usepackage{circuitikz}
\usepackage{enumitem}
\usepackage{multirow}
\usepackage{pgf}
\usepackage{dashrule}
\usepackage{tcolorbox}
\usepackage{booktabs}
\usepackage{makecell}
\usepackage{placeins}

\usepackage{multicol}
\usepackage{hyperref}
\usepackage{cleveref} 
\usepackage{xcolor}
\usepackage{ulem}

\definecolor{LinkBlue}{HTML}{00AEEF}
\hypersetup{
colorlinks=true,
linkcolor=LinkBlue,
citecolor=LinkBlue,
urlcolor=LinkBlue
}



\def\tsc#1{\csdef{#1}{\textsc{\lowercase{#1}}\xspace}}
\tsc{WGM}
\tsc{QE}
\tsc{EP}
\tsc{PMS}
\tsc{BEC}
\tsc{DE}

\begin{document}
\bibpunct{[}{]}{,}{n}{}{,}
\shorttitle{Hybrid dynamical systems modeling of power systems}
\shortauthors{B.G. Odunlami et~al.}
\title [mode = title]{Hybrid dynamical systems modeling of power systems} 
\author[aff1]{Bukunmi Gabriel Odunlami}
\cormark[1]
\ead{bgo@njit.edu}
\author[aff1]{Marcos Netto}
\affiliation[aff1]{organization={Department of Electrical and Computer Engineering, New Jersey Institute of Technology},
city={Newark},
postcode={07102}, 
state={New Jersey},
country={USA}}
\cortext[cor1]{Corresponding author}

\author[aff2]{Yoshihiko Susuki}
\affiliation[aff2]{organization={Department of Electrical Engineering, Kyoto University},
city={Kyoto},
postcode={615-8510},
country={Japan}}

\begin{abstract}
Hybrid dynamical systems offer a rigorous foundation for representing mixed continuous–discrete dynamics and have become increasingly relevant in the analysis of modern power systems. Despite their potential, existing studies remain focused on isolated applications or case-specific implementations, offering limited generalizability and guidance for model selection. This paper addresses that gap by providing a comprehensive overview of hybrid modeling approaches relevant to power systems. It critically examines key formalisms, including hybrid automata, switched systems, and piecewise affine models, evaluating their respective strengths, limitations, and suitability across system analysis, stability assessment, and control design tasks. Quantitative comparisons show that hybrid automata and switched-system models achieve substantially lower modeling error than mixed-logical dynamical abstractions in nonlinear hybrid scenarios, and that explicit hybrid formulations improve state-estimation accuracy by one to two orders of magnitude compared to continuous smoothing approaches. The review further consolidates these insights into a structured comparison that clarifies how different hybrid modeling formalisms map to specific analysis and control tasks in renewable-rich, converter-dominated grids. In doing so, the paper identifies open challenges and outlines future research directions to support the systematic application of hybrid systems modeling and methods in modern power systems.
\end{abstract}


\begin{keywords}
  hybrid automata \sep hybrid dynamical systems \sep mixed logical dynamical systems \sep power systems modeling \sep renewable energy systems \sep switched systems
\end{keywords}

\maketitle

\section{Introduction} \label{sec:intro}
Electric grids are undergoing a profound transformation driven in part by the integration of renewable energy sources. While renewable energy technologies support decarbonization and energy sustainability goals, they also introduce faster switching dynamics, low-inertia, operational uncertainty, and new control-driven instabilities. These characteristics have led to more volatile power flows, reduced stability margins, and a growing number of system events driven by discrete or logic-based operations. 

Power system dynamics are commonly represented using continuous-time models, which assume that system variables evolve smoothly. However, modern power systems are increasingly influenced by discrete events resulting from internal operations or external disturbances.  Such events include the sequential tripping of power system components \cite{9733408}, microgrid islanding and reconnection to the grid \cite{10467393}, and fault clearance by protective devices \cite{7741821}. Other examples and their implications in renewable-rich systems are summarized in \autoref{tab:significant discrete events}.

To account for such discrete phenomena, continuous models introduce approximations or simplifying assumptions which may obscure critical transient behaviors or yield flawed assessments of stability, protection coordination, and control performance \cite{YOUNESI2022112397}. As power networks evolve into decentralized, converter-dominated architectures with increasing dependence on digital control and communication layers, continuous-time differential–algebraic equations (DAE) alone are insufficient; switching actions and discrete events must be systematically incorporated as they can substantially alter system trajectories \cite{tOWARDSRESILIENT,9893118}.

 Hybrid dynamical system (HDS) theory provides a unified framework for representing such mixed continuous–discrete behavior. HDS methods have gained traction in several engineering domains and are increasingly relevant in power systems with high levels of renewable integration. Additionally, they have been used to model cyber-physical interdependencies between network behavior and discrete cyber processes. Applications in this area include addressing cybersecurity challenges such as false data injection attacks, switching attack events, and the impact of communication delays on the performance of control systems \cite{Cyberpower}.

HDS theory has roots in non-smooth dynamical systems and control theory. The field began to take formal shape with major contributions such as the introduction of mixed-logic systems \cite{Witsenhausen1966} and Filippov’s differential inclusions for discontinuous vector fields \cite{Filippov1988}, followed by advances in hybrid optimal control \cite{Branicky1998, antasaklis1995hybrid}. Early applications in power systems appeared in system analysis \cite{hikihara2005} and limit-cycle stability assessment \cite{980200}.

Despite these advances, the adoption of HDS frameworks in power system modeling remains relatively modest. A clearer understanding is needed of when hybrid formulations offer tangible advantages over smooth continuous-time models, how different HDS frameworks compare, and the extent to which they provide more faithful representations of the event-driven and mode-dependent behaviors in modern power systems.

This review addresses these gaps by providing a structured overview of hybrid system methodologies and their applications in power systems. The main contributions of this work are threefold:

\begin{enumerate}[nosep]
\item A comprehensive review of HDS applications in power systems with high penetration of renewable energy resources. This includes diving into broad research categories such as modeling, analysis, control, and design.
\item A structured guidance on selecting an HDS modeling framework, e.g., hybrid automata, switched systems, piecewise affine models, among others, while evaluating their relative strengths, limitations, and suitability for distinct classes of power system problems.
\item An identification and discussion of key research gaps, highlighting methodological limitations and opportunities for future research in HDS modeling of modern power systems.
\end{enumerate}

\begin{table*}[!t]
\scriptsize
\setlength{\tabcolsep}{3pt}
\renewcommand{\arraystretch}{1.05}
\begin{tcolorbox}[colframe=black, colback=white, boxrule=0.6pt, width=\textwidth]

\noindent
\begin{minipage}[t]{0.49\textwidth}
\textbf{Nomenclature}
\end{minipage}\hfill
\begin{minipage}[t]{0.49\textwidth}
\raggedleft\textbf{Abbreviations}
\end{minipage}
\vspace{0.3em}

\noindent
\begin{minipage}[t]{0.49\textwidth}
\raggedright
\begin{tabular}{@{}l l@{}}
\textit{A\textsubscript{t}} & System matrix in MLD model \\
\textit{B\textsubscript{t}} & Input matrix in MLD model \\
\textit{C} & Flow set \\
\textit{C\textsubscript{t}} & Output matrix in MLD model \\
\textit{D} & Jump set \\
\textit{D\textsubscript{t}} & Feedforward matrix in MLD model \\
\textit{E\textsubscript{1t}-E\textsubscript{5t}} & Disturbance matrix in MLD model \\
\textit{E} & Set of edges \\
\textit{F(x)} & Flow map \\
\textit{G} & Guard condition \\
\textit{G(x)} & Jump map \\
\textit{$H_1, H_2$} & Individual hybrid automaton in HIOA \\
\textit{q} & Discrete state of HA \\
\textit{R} & Reset map \\
\textit{$\sigma(t)$} & Switching signal \\
\textit{S\textsubscript{0}} & Set of initial states \\
\textit{S\textsubscript{f}} & Set of final states \\
\textit{$\zeta(t)$} & Flat state in differential flatness \\
\textit{x(t)} & State vector \\
\textit{x\textsuperscript{+}} & State after a jump \\
\textit{u(t)} & Control input vector \\
\textit{$\mathbb{R}^n$} & \( n \)-dimensional real space \\
\textit{$\Phi(z, \dot{z}, \ddot{z}, \dots)$} & State map in differential flatness \\
\textit{$\Psi(z, \dot{z}, \ddot{z}, \dots)$} & Input map in differential flatness \\
\textit{$U_1, U_2$} & Input sets in HIOA \\
\textit{$X_1, X_2$} & A set of states in HIOA \\
\textit{$\chi^2$} & Chi-square distribution \\
\textit{$V$} & Admissible input set for ($q,q'$) \\
\textit{$Y_1, Y_2$} & Output sets in HIOA \\
\end{tabular}
\end{minipage}\hfill
\begin{minipage}[t]{0.46\textwidth}
\raggedright
\hspace{0.02\textwidth} 
\begin{tabular}{@{}l l@{}}
AVR & Automatic voltage regulator \\
CGHDS & Controlled general hybrid dynamical system\\
DAD & Differential-algebraic discrete \\
DAE & Differential-algebraic equations \\
DAIS & Differential-algebraic impulsive system \\
DHA & Discrete-hybrid automata \\
DMIB & Double machine infinite bus \\
DSAR & Differential, switched algebraic and state-reset \\
FFNN & Feed forward neural network \\
FHA & Flat hybrid automata \\
GFL & Grid following\\
GFM & Grid forming \\
HA & Hybrid automata \\
HDS & Hybrid dynamical system \\
HIOA & Hybrid input/output automata \\
HPN & Hybrid Petri net \\
MIDP & Mixed-integer dynamic programming\\
MILP & Mixed-integer linear programming \\
MIQP & Mixed-integer quadratic programming \\
MLD & Mixed logic dynamical \\
MMPS & Multimachine power system \\
MPC & Model predictive control \\
PWA & Piece-wise affine \\
SHA & Stochastic hybrid automata \\
SMIB & Single machine infinite bus \\
SVC & Static var compensator \\
VPP & Virtual power plant \\
WTG & Wind turbine generator \\
\end{tabular}
\end{minipage}
\end{tcolorbox}
\normalsize
\end{table*}
\FloatBarrier

This paper is not intended as an exhaustive review of all HDS methods, instead, it focuses on the key frameworks predominant in power system applications and the challenges introduced by converter-based, event-driven, and cyber–physical dynamics.
\begin{table*}[!t]
\caption{Examples of discrete events in modern power systems with renewables}
\centering
\begin{tabular}{p{3.5cm}p{5cm}p{5.4cm}p{1.5cm}}
\toprule
\textbf{Event type} & \textbf{Root cause} & \textbf{Implication} & \textbf{Reference} \\
\midrule
Load shedding & Triggered by frequency drops due to renewable variability & Instability if not coordinated with inverter controls & \cite{9733408} \\
On-load tap changer & Responds to voltage deviations from solar photovoltaic generation & Discrete tap changes introduce delays and jumps & \cite{1397735} \\
Capacitor bank switching & Counteract reactive power imbalance from renewable energy resources & Sudden reactive injection may cause voltage oscillations & \cite{Fink1999} \\
Unit commitment & Adjusted frequently due to uncertain renewable energy resources forecasts & Introduces discrete operational shifts and commitment delays & \cite{8789685} \\
Inverter mode transitions & Switch between normal, fault-ride through, or droop modes & Changes system dynamics and stability margins & \cite{4237906} \\
\makecell[l]{Microgrid islanding/\\ reconnection} &
\makecell[l]{Triggered by faults in renewable \\ energy resources clusters} &
\makecell[l]{Alters grid topology with potential \\ synchronization issues} &
\cite{islanding} \\
\bottomrule
\end{tabular}
\label{tab:significant discrete events}
\end{table*}
The rest of this paper is organized as follows: \autoref{sec:hds_power} explores mathematical frameworks in HDS theory, illustrating their applications in renewable energy and power systems. \autoref{sec:verification} discusses safety verification methods and the concept of reachability analysis in the context of hybrid power system modeling. \autoref{sec:applications} presents a comprehensive review of the use of HDS models in power systems, categorizing specific applications by the overarching problems they address and providing insights into the current state of research, while identifying potential research directions. \autoref{sec:discussion} summarizes the state-of-the-art developments and highlights additional research opportunities. \autoref{sec:conclusion} concludes the paper.

\begin{figure*}[!t]
\centering
\includegraphics[width=0.73\linewidth]{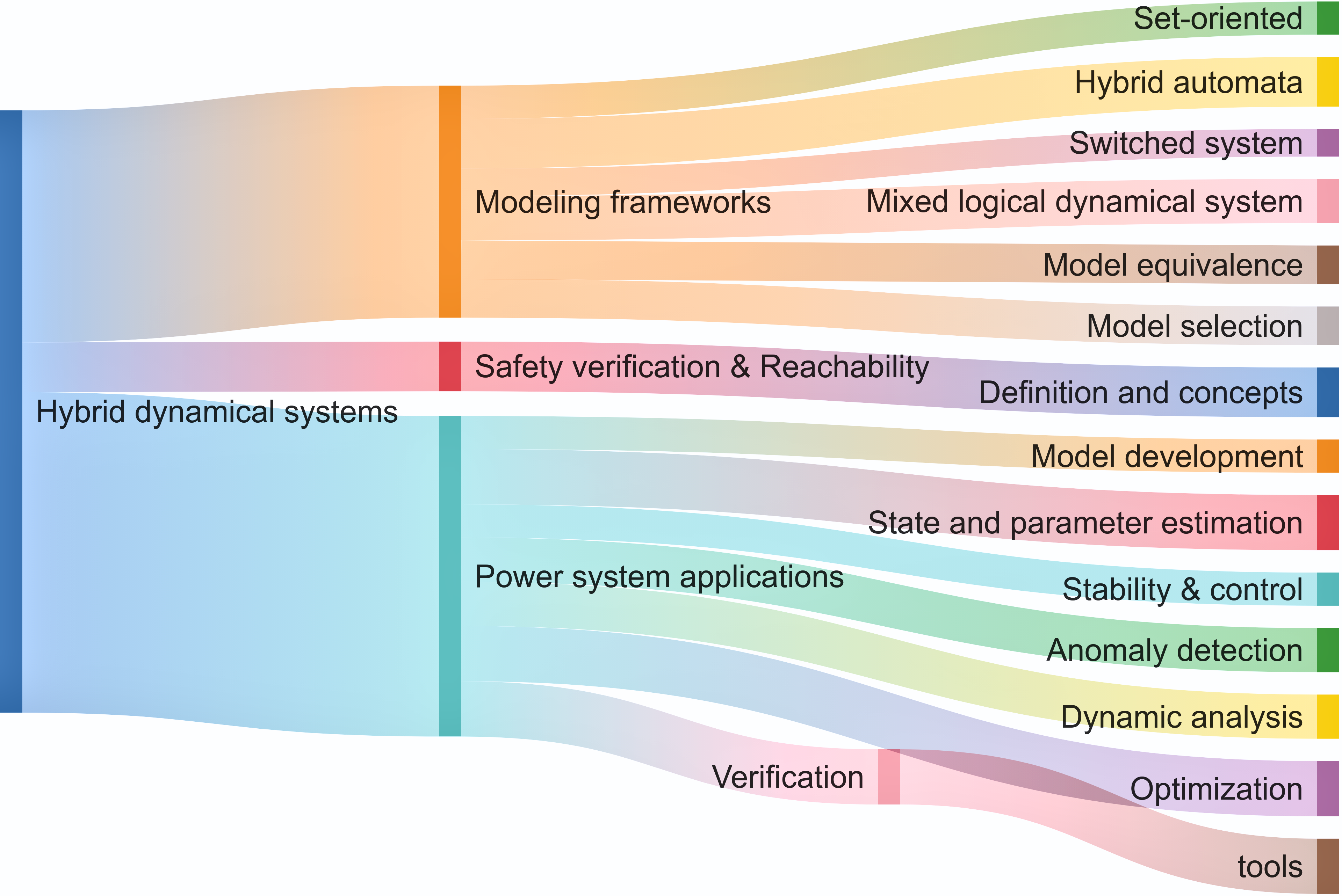}
\caption{Overview of proposed review and hybrid dynamical systems topics.}
\label{fig:discrete_events_res}
\end{figure*}

\section{Overview of HDS modeling frameworks} \label{sec:hds_power}
This section presents a concise overview of foundational HDS formalisms that have found relevance in converter-dominated and renewable-rich power systems. The goal is to enhance understanding of how these frameworks are structured and how they may be applied in modern power system modeling. We begin by characterizing the structural behavior of general hybrid systems, followed by a focused summary of each modeling approach.

In this paper, the term hybrid system is used in accordance with Definition 1, and should not be confused with broader uses of "hybrid" that imply combining different methods, models, or technologies.

\textbf{Definition 1.} (Hybrid dynamical system): A hybrid dynamical system is one that exhibits both continuous and discrete dynamics—characterized by differential or difference equations governing continuous behavior and discrete transitions triggered by logic rules, thresholds, event conditions, or switching mechanisms.

Hybrid systems can exhibit a variety of structural behaviors. In some cases, transitions between discrete modes may induce changes to the system dynamics, parameters, or state space. In other cases, discrete events occur without altering the underlying continuous dynamics. For instance, consider a generator delivering power through two parallel transmission lines with identical electrical properties. A maintenance-driven switch from one line to the other constitutes a discrete mode change. In principle, the system's continuous dynamics, such as the voltage at the generator's terminal or the power delivered to the load, remain the same before and after the switching.

\begin{figure*}[!t]
\centering
\includegraphics[
  width=0.79\linewidth
]{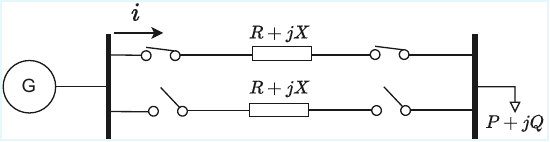}
\caption{Illustrative example: Single generator–load bus with switching between two identical transmission lines.}
\label{fig:illustrative-example}
\end{figure*}
Yet, such transitions often coexist with protection schemes and stochastic phenomena, including uncertainties in switching times and event triggers, which may lead to non-deterministic transitions and ambiguous mode definitions. Accordingly, in the following subsections, we discuss prominent hybrid system modeling approaches, including mixed logical dynamical (MLD) systems, hybrid automata (HA), switched systems, and set-oriented methods. For each, we summarize its mathematical foundations and suitability within power system applications.
\subsection{Set-oriented modeling framework} \label{subsec:set_oriented}
A hybrid system, typically denoted as \(\mathcal{H} \), is represented by data components \((C, f, D, g)\) \cite{goebel2012hybrid}. These components define the system's behavior through a combination of continuous flow and discrete jumps. A hybrid system is compactly described by~\eqref{eq:Dataset}.

\begin{equation}
\mathcal{H} =
\begin{cases}
\dot{x} = f(x), & \text{if } x \in C,\\
x^{+} = g(x), & \text{if } x \in D.
\end{cases}
\label{eq:Dataset}
\end{equation}

In this formulation, \( x \in \mathbb{R}^n \) represents the state variables, the function \( f : C \rightarrow \mathbb{R}^n \) is the flow map defining the continuous evolution of the system within the flow set \( C \subseteq \mathbb{R}^n \). The jump map \( g : D \rightarrow \mathbb{R}^n \) dictates the system's behavior during discrete events, which occur when the state is within the jump set \( D \subseteq \mathbb{R}^n \), and \( x^+ \) denotes the state immediately after such a discrete event. The first derivative of the state variables satisfies the flow condition.
\FloatBarrier
A solution to \eqref{eq:Dataset} is represented by a hybrid arc, parameterized over $(t,j) \in \mathbb{R}_{\geq 0} \times \mathbb{N}$, where $t$ tracks ordinary time and $j$ counts discrete transitions. The hybrid time domain is thus constructed from a non-decreasing sequence of time intervals $\{t_j\}_{j=0}^{J+1}$ ensuring well-defined flows and jumps. This guarantees that such systems evolve coherently: between any two jumps $j$ and $j+1$, the system undergoes continuous-time evolution over $[t_j, t_{j+1}]$, and a jump may occur only at $t_{j+1}$. 

\subsubsection{Illustrative example} \label{subsubsec:example1}
Consider the scenario in \autoref{fig:illustrative-example} where the generator is a single synchronous machine, described by:
\begin{equation}
\begin{aligned}
\dot{\delta} &= \omega \\
M \dot{\omega} &= P_m - P_e - D\omega
\end{aligned}
\label{eq:swing_equation}
\end{equation}
where \( \delta \) is the rotor angle, \( \omega \) is the speed deviation, \( M \) is the inertia constant, \( D \) is the damping coefficient, and \( P_e \) is the electrical power transferred through the transmission line. While we understand the assumptions behind \eqref{eq:swing_equation} \cite{Sauer1997}, this simple model serves the purpose of this example.

Suppose that when the current in the active transmission line exceeds a preset protection threshold $I_{max}$, the line is disconnected and power is immediately rerouted to the other line. Once the offline line is restored to service, it resumes carrying power according to the prevailing operating conditions. Using \eqref{eq:Dataset}, the hybrid dynamics can be expressed as:
\begin{align}
\mathcal{H} =
\begin{cases}
\dot{x} = f_1(x), & x \in C_1 \quad \text{(line 1 Active)} \\
\dot{x} = f_2(x), & x \in C_2 \quad \text{(line 2 Active)} \\
x^+ = g(x), & x \in D \quad \text{(switch condition)}
\end{cases}
\label{set-orientedeq}
\end{align}
where \( x = \begin{bmatrix} \delta & \omega \end{bmatrix}^T \in \mathbb{R}^2\). The functions \( f_1(x) \) and \( f_2(x) \) describe the continuous dynamics under each transmission line. Since both transmission lines have identical characteristics, the continuous dynamics simplify to:
\begin{equation}
f_1(x) = f_2(x) =
\begin{bmatrix}
\dot{\delta} \\
\dot{\omega}
\end{bmatrix}
=
\begin{bmatrix}
\omega \\
\frac{1}{M} \left( P_m - P_e - D\omega \right)
\end{bmatrix}
\end{equation}

\noindent The flow set for which each transmission line remains active can be described by $C_1 \cup \ C_2$, with
\begin{equation}
\begin{aligned}
C_1 &= \{ x \mid I_{\ell_{1}} < I_{\max} \} \\
C_2 &= \{ x \mid P_{e,\,\ell_{1}} \notin (P_{\min}, P_{\max}) \}
\end{aligned}
\end{equation}
where \( I_{\ell_{1}} \) is the current flowing through line 1, \( I_{\max} \) is the overload threshold current, and \( P_{e,\,\ell_{1}} \) is the active power transferred through line 1. The jump set is defined as follows:
\begin{equation}
D_1 \cup D_2 = \{ x \mid I_{\ell_{1}} > I_{\max} \} \cup \{ x \mid P_{\min} \leq P_{e,\,\ell_{1}} \leq P_{\max} \}
\end{equation}
Given that the state variables are not updated as the transmission lines are switched,
\begin{equation}
g(x) =
\begin{cases}
q_2, & \text{if } I_{\ell_{1}} > I_{\max} ) \\
q_1, & \text{if } P_{\min} \leq P_{e,\,\ell_{1}} \leq P_{\max} )
\end{cases}
\label{set_orientedeq2}
\end{equation}
where $q_1$ and $q_2$ are active lines 1 and 2, respectively.

\subsubsection{Set-valued generalization for uncertain dynamics}
The hybrid system \eqref{eq:Dataset} can be extended to a differential inclusion framework by replacing the single-valued flow and jump maps with set-valued mappings such that $\dot{x} \in F(x)$ and $x^+ \in G(x)$ where \( F: \mathbb{R}^n \rightrightarrows \mathbb{R}^n \) is a set-valued map defining the possible trajectories of the system within the flow set \( C \), and \( G: \mathbb{R}^n \rightrightarrows \mathbb{R}^n \) describes the possible state transitions when \( x \in D \) \cite{goebel2012hybrid}.

\textbf{Definition 2.} (Differential inclusion): A differential inclusion generalizes classical differential equations by allowing the derivative of the state to belong to a set of possible values rather than being uniquely determined. 

This generalization to differential inclusions makes the framework particularly useful in converter-dominated power systems, where switching events, control constraints, and operational uncertainties often prevent the definition of a unique flow function. The flexibility to model system evolution within a set of admissible dynamics enables a more faithful representation of complex behaviors such as ambiguities in switching due to protection logic or hysteresis, nondeterministic transitions induced by software-based controls or communication delays, and uncertainties arising from converter responses, power generation intermittency, and load fluctuations.

In such a context, classical ordinary differential equation-based formulations may be inadequate to represent the full range of behaviors emerging from uncertain transitions and structural ambiguities. In contrast, differential inclusion frameworks enable the modeling of:
\begin{itemize}[nosep, leftmargin=.5cm]
\item Uncertain or time-varying control responses, as observed in droop-based or virtual inertia schemes.
\item Non-deterministic state evolution resulting from ambiguous switching conditions or event-triggered logic.
\item Complex interactions among dynamic constraints, such as current and voltage limits, or state-of-charge bounds in energy storage systems.
\end{itemize}

For example, \cite{reddy2023decentralized} formulates a hybrid system model of shipboard power systems with integrated energy storage and renewable generation, leveraging the flexibility of differential inclusions to examine switching transients and coordinated power flows. Similarly, modeling and designing power systems from a cyber-physical systems perspective \cite{sanfelice2011cyber} can utilize this framework, where the grid's continuous dynamics develop alongside discrete events from controllers, relays, and communication networks.

Despite their theoretical appeal, this framework \eqref{eq:Dataset} remains relatively underexplored in power systems. This is further discussed in \autoref{sec:discussion}.

\subsection{Hybrid automaton modeling framework} \label{subsec:ha_model}
Following \cite{lin2018hybrid, Henzinger95}, a hybrid automaton is described by a collection 
\begin{equation}
\mathcal{H} = \{Q, X, f, \text{Init}, \text{Inv}, E, G, R\}
\label{eq: HA}
\end{equation}
where \( Q = \{q_1, q_2, \ldots\} \) is a finite set of discrete states, representing the different modes or configurations of the system. \( X \subseteq \mathbb{R}^n \) represents the state space of the continuous state variables. \( f: Q \times X \to \mathbb{R}^n \) governs the continuous dynamics within each discrete state. \( \text{Init} \subseteq Q \times X \) is the set of initial states, specifying the starting discrete state and the initial values of the continuous variables. \( \text{Inv}: Q \to 2^X \) assigns to each discrete state \( q \in Q \), an invariant set \( \text{Inv}(q) \subseteq X \) where the system must remain within while in the discrete state \( q \). \( E \subseteq Q \times Q \) is the set of discrete transitions. \( G: E \to 2^X \) assigns to each discrete transition \( (q, q') \in E \), a guard set \( G(q, q') \subseteq X \) where a transition can only occur if the continuous state is within the guard set. \( R: E \times X \to 2^X \) is the reset map specifying how the continuous state variables are updated when a discrete transition occurs.

In \eqref{eq: HA}, continuous inputs, disturbances, and discrete disturbance symbols are not considered for simplicity and clarity. These aspects are often necessary for solving control problems in power grid dynamics \cite{takatsuji2009hybrid, takatsuji2011hybrid}. The trajectories of $\mathcal{H}$ evolve both continuously and through discrete jumps. A detailed mathematical description of these trajectories and their behavior can be found in \cite{Tomlin2003, PYTLAK2022101202} and the references therein.

\begin{figure*}[ht!]
\centering
\begin{tikzpicture}[->, >=stealth', auto, semithick, node distance=3cm, scale=0.92, every node/.style={scale=0.85}]
\clip(-3.55, -1.9) rectangle (6.9, 1.2);
\node[state, minimum size=1.2cm, align=center] (q1) {\footnotesize q1 \\ $f(q1, X)$ \\ $R(q1, X)$};
\node[state, minimum size=1.2cm, align=center, right of=q1] (q2) {\footnotesize q2 \\ $f(q2, X)$ \\ $R(q2, X)$};
\path (q1) edge [loop left] node {\footnotesize G(q1, q1)} (q1)
            edge [bend left] node {\footnotesize G(q1, q2)} (q2);

\path (q2) edge [loop right] node {\footnotesize G(q2, q2)} (q2)
            edge [bend left] node {\footnotesize G(q2, q1)} (q1);
\end{tikzpicture}

\vspace{-0.6cm}
\caption{Hybrid automata model.}
\label{fig:hybrid_automata}
\end{figure*}
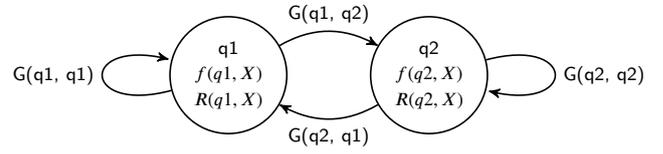

HA extends finite automata by incorporating continuous dynamics within each discrete mode. Each discrete state has a corresponding continuous state, which is governed by specific DAE. Events trigger transitions between discrete states as depicted in \autoref{fig:hybrid_automata}. These transitions can introduce discontinuities in algebraic variables and may cause the state variables to reset or the continuous state space to change.
The HA model has been widely adopted in power system studies, supporting analyses of transient stability \cite{Susuki2012}, voltage stability \cite{susuki2007predicting}, and operational challenges in renewable-rich grids. Applications span energy management \cite{kafetzis2020energy}, microgrid control \cite{doi:10.1049/iet-rpg.2019.0664}, control design for grid-connected solar photovoltaic systems \cite{HekssPv}, and building energy optimization \cite{oberkirsch2020}.

\subsection{Switched systems modeling framework} \label{subsec:switched_systems}
A switched system is a particular case of a hybrid system characterized by a family of subsystems and a rule that governs switching among them. It can be expressed as:
\begin{equation}
\dot{x}(t) = f_{\sigma(t)}(x(t), u(t))
\label{eq:switched}
\end{equation}
where \( \sigma(t) \in \mathcal{Q} = \{1, 2, \ldots, N\} \) is the switching signal selecting the active mode at time \( t \), and \( f_i(\cdot) \) defines the dynamics of the $i$th subsystem. Each mode may correspond to distinct converter controls, operating states (e.g., islanding or grid-following), or resource management logic, such as battery charging or generator dispatch.

A central challenge in such systems is understanding how frequent mode transitions triggered by protection schemes, control logic, or renewable intermittency impact overall system stability. As shown in \cite{liberzon1999stability, zhai2007note}, the properties of the switching signal, such as dwell time or mode sequences, critically influence the system's behavior.

To account for all possible trajectories induced by switching events in a large network, \eqref{eq:switched} can be rewritten in the framework of differential inclusion. Using \eqref{eq:Dataset}, we have \cite{goebel2012hybrid}:
\small
\begin{equation}
\dot{x} = 
\begin{cases} 
\begin{pmatrix} f_q(z) \\ 0 \end{pmatrix} & x \in C \\
\begin{pmatrix} z \\ q^+ \end{pmatrix} & x \in D
\end{cases}
\end{equation}
\normalsize
where \( x = \begin{pmatrix} z \\ q \end{pmatrix} \), with \( z \in \mathbb{R}^n \) as the continuous state and \( q \in Q \) as the discrete mode. This is useful when dealing with a switched system as a subsystem in a large application modeled by \eqref{eq:Dataset}.

Switched system models are particularly fitting for stability analysis in modern power systems, as explored in \cite{chen}. For example, a switched system approach enables structured analysis of how inverter control modes adapt to varying grid conditions or how backup sources take over during intermittent generation resources dropouts \cite{tan2021}.

Beyond stability, switched systems support optimization-based coordination of distributed energy resources. In \cite{wu2018}, the economic dispatch problem in microgrids is modeled as an optimal control problem over switching dynamics, where the timing and sequencing of power source transitions (e.g., diesel generators, batteries, renewable energy sources) are decision variables. To address the complexity of determining optimal switching instants, the authors employ a time-scaling transformation and introduce penalty functions to enforce minimum dwell-time constraints. This approach not only improves operational efficiency and reduces fuel costs but also mitigates excessive switching, which ultimately prolongs battery life and reduces mechanical stress on generators.

\subsection{Mixed logical dynamical systems modeling framework}
The MLD framework represents a control-centric subclass of hybrid systems, where continuous dynamics are augmented with logic-based constraints. It models a system behavior using linear equations that combine real-valued and binary variables, subject to linear inequality constraints. Following \cite{lin2018hybrid}, the general form is:
\begin{align}
x^+ &= A_t x + B_{1t} u + B_{2t} \delta + B_{3t} z \label{eq:MLD_state}\\
y &= C_t x + D_{1t} u + D_{2t} \delta + D_{3t} z \\
\text{s.t. } & E_{2t} \delta + E_{3t} z \leq E_{1t} u + E_{4t} x + E_{5t}
\label{eq:MLD}
\end{align}
where \(x\), \(u\), and \(y\) respectively represent the state, input, and output vectors. The vector of binary variables \(\delta\) encode switching decisions or logic rules, while \(z\) includes auxiliary variables introduced during logic-to-inequality transformations.

This formulation enables logic-based decisions to be embedded directly into state-update equations. Thus, it provides a unified structure for casting control problems as mixed-integer optimization problems, which is especially advantageous when deployed with model predictive control (MPC). For example, \cite{lian2017mixed} demonstrates its use for predictive energy management in plug-in hybrid electric vehicles, where mode-switching logic and power constraints are co-optimized to reduce fuel consumption. This logic-based optimization approach is also extended to residential and microgrid applications. For instance, \cite{4919363} models residential energy systems (rooftop solar photovoltaics, battery storage, combined heat and power) using the MLD approach, replacing embedded on/off logic with linear constraints to enable flexible decision-making under operational and economic limits. The resulting MPC optimizes storage, generation, and power flows, thereby improving self-consumption, reducing costs, and enhancing resource utilization. In a related study, \cite{HybridMPC} proposes a DC microgrid formulation that incorporates generation (renewable energy source and otherwise) and storage units, utilizing the DC bus energy as a state variable and integrating continuous dynamics with logic constraints to mitigate voltage fluctuations and minimize unnecessary battery switching.

These examples demonstrate how the MLD framework supports scalable and optimization-ready modeling of complex systems. Its ability to represent both logical rules and continuous physics in a single formalism makes it particularly compelling for energy management strategies involving operational switching and constrained multi-energy coordination.

\subsection{Model equivalence and other relevant frameworks} \label{subsec:model_equivalence}
Other hybrid modeling formalisms, such as piecewise affine (PWA) systems and hybrid Petri nets (HPN), have also found application in power system, including renewable energy integration \cite{hybridPetrinet, windpower, SindarehWindEnergy} and microgrid coordination \cite{10.4018/IJAMC.2020010105}. Some of these can be interpreted as particular cases or extensions of the core frameworks discussed earlier. For example, a PWA system can be viewed as a switched system where the dynamics transition between affine subsystems is based on the state’s position relative to a set of linear hyperplanes partitioning the state space \cite{lin2018hybrid}. This provides a flexible means of modeling mode-dependent behaviors or nonlinearities through linear approximations. The interested reader is referred to \cite{CABRAL2023556,8123285} for details of these models.

Interestingly, these models can be equivalent under certain conditions—such as well-posedness, boundedness of input, state, and output variables, as well as discretization and inequality approximations \cite{heemels2001equivalence}. For example, any well-posed PWA system can be transformed into an MLD system and vice versa. Indeed, \cite{heemels2001equivalence} demonstrated this using five modeling frameworks, although the author acknowledges that their approach makes mild assumptions. 

Since choosing a modeling framework is case-dependent, model equivalence becomes significant, as it allows the transfer of analysis techniques, control design methods, and verification tools between different models. Practically, it means that one can choose the modeling framework that best suits the analysis or task at hand, knowing that the results will be valid across equivalent models.

It is also worth noting that there could be combinations of multiple modeling frameworks to tackle complex power systems problems. Furthermore, machine learning techniques have been incorporated, including neural-network-based HA \cite{yang2023data} and probabilistic transition rules, as seen in stochastic HA \cite{erbes2005stochastic}.

\subsection{Model selection of HDS methods in power systems}
System requirements, modeling objectives, and computational feasibility influence the choice of a hybrid modeling approach.

MLD models are particularly advantageous for control and optimization problems, making them suitable for systems where decision-making under constraints is crucial. Their compatibility with optimization algorithms, such as mixed-integer linear programming (MILP), enhances their tractability in complex control tasks \cite{inbookModelHybrid}. A key drawback of this framework is its assumption of linear system dynamics and constraints, which may not always be applicable in power systems.

Conversely, switched systems and PWA models excel in capturing nonlinear dynamics and state-dependent switching behaviors. PWA models, in particular, can prevent mathematical artifacts like Zeno behavior which can be described informally as the system making infinitely many jumps in a finite amount of time by enforcing piecewise affine transitions that stop infinite switching within finite time ~\cite{BakoYahya2019}. This ensures physically meaningful transitions between control modes. However, the PWA model assumes that the system can be approximated by a set of affine dynamics over predefined regions, which can be challenging to implement in large-scale power networks due to nonlinearities caused by factors such as saturation limiters and deadbands. Power systems generally violate the assumption of piecewise linearity, making accurate PWA abstraction challenging. For this reason, the unabstracted system-level behavior is more faithfully represented using switched system models, which naturally subsume PWA as a special case but allow for broader nonlinear mode-dependent dynamics without requiring explicit affine partitioning.

HA are well-suited for modeling event-driven systems where discrete interactions significantly govern system evolution. Their formulation is especially intuitive when applied to systems with linear guard conditions and reset maps that define how state variables are updated upon entering a new mode. Similar to HA, HPN is effective in modeling systems where the timing and sequencing of events are critical. For instance, coordinating energy flow in a renewable energy system where solar, wind, and battery components must interact based on availability, demand, and priority constraints.

Set-oriented methods aim to generalize and unify multiple hybrid modeling paradigms within a single mathematical framework. They offer a high degree of abstraction and can encapsulate uncertainty and non-determinism. 

Next, we compare these hybrid modeling frameworks using the following four qualitative criteria: ability to handle nonlinearities, switching mechanism, uncertainty handling, and flexibility, as shown in \autoref{tab:hybrid_model_comparison}. Additionally, for each framework, we suggest power system applications where the structural strengths are organically suited or commonly applied due to tooling support.
\begin{table*}[h!]
\scriptsize
\centering
\caption{Qualitative comparison of HDS frameworks for renewable-rich power systems}
\label{tab:hybrid_model_comparison}
\begin{tabular}{@{}p{1.7cm}p{1.3cm}p{1.5cm}p{1.4cm}p{3.2cm}p{5cm}@{}}
\toprule
\textbf{Model} & \textbf{Nonlinear or linear} & \textbf{Switching mechanism} & \textbf{Uncertainty handling} & \textbf{Flexibility} & \textbf{Suitable power system applications} \\
\midrule

MLD & Linear & Logic with linear inequalities & Deterministic logic only & Supports logic-rich control but within a linear, solver-constrained form & Predictive control, logic-based coordination, optimization tasks \\
 
PWA & Locally linear segments & Region-bound transitions (state-based) & Deterministic event only & Rigid at boundary crossings & Local controller design, model-based identification, inverter mode partitioning \\

HA & Generally nonlinear & Guard-triggered transitions (time/state) & Probabilistic extensions in components & Highly expressive: supports discrete resets and mode extensions & Formal verification, supervisory control, state and parameter estimation, protection schemes, stability analysis \\

Set-oriented & Generally nonlinear & Guard-triggered transitions (time/state) & Set-valued differential inclusion & Modular and set-extendable & Resilience assessment, safety verification, uncertainty-aware modeling \\

Switched & Subsystem-specific (often nonlinear) & Predefined/rule-driven & Probabilistic switching rule only & Constrained by predefined models & Structured mode scheduling(e.g. GFM/GFL cycling), stability analysis \\
\bottomrule
\end{tabular}
\end{table*}
\normalsize

\textit{Remarks:} While each framework naturally lends itself to a particular task, all models are theoretically capable of being extended with varying degrees of abstraction or complexity to cover a wide range of hybrid system applications. The association listed above reflects what is naturally suited, rather than being exclusive or exhaustive.

\begin{figure*}[ht]
    \centering
    \includegraphics[width=0.86\linewidth]{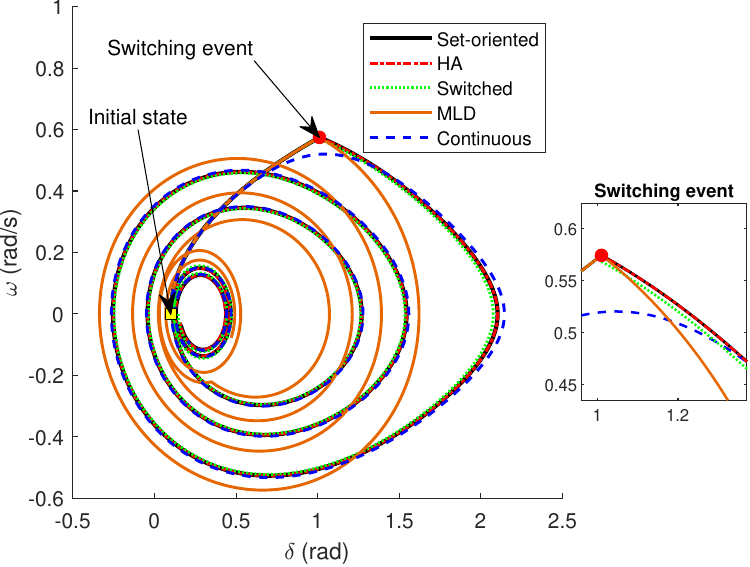}
    \caption{Phase portrait of a single generator–load bus system with switching between two identical transmission lines using five modeling formalisms.}
    \label{phase_portrait}
\end{figure*}

To further illustrate the modeling implications of the discussed 
formalisms, we evaluate five approaches on the nonlinear hybrid 
scenario described in \autoref{subsubsec:example1}. From \autoref{phase_portrait}, the set-oriented 
reference in \eqref{set-orientedeq}--\eqref{set_orientedeq2}, HA and 
switched models, produce nearly identical 
trajectories because the hybrid evolution is governed by a 
piecewise-constant mode schedule and the reset map is trivial 
($x^{+}=x$). In such cases, a continuous approximation that smooths the 
mode transitions remains adequate. In contrast, the MLD representation exhibits significantly 
larger modelling error due to its affine embedding of the nonlinear 
restoring torque. In particular, the electrical power term $P_e=b\sin\delta$ must be replaced 
by a first-order linear approximation.

The error metrics in \autoref{tab:SMIB} quantify the deviation of each modeling 
framework from the set-oriented reference. The RMSE values measure the 
average trajectory discrepancy in rotor angle~$\delta$ and speed~$\omega$, 
while Max indicate the 
largest instantaneous deviation over the simulation horizon. The HA and switched models yield the smallest error metrics, followed closely by the continuous approximation. The MLD model, while computationally fast, exhibit the largest discrepancy due to the accumulated linearization error.
\begin{table*}[ht]
\centering
\caption{Modeling error metrics relative to the set-oriented reference.}
\begin{tabular}{lccccc}
\toprule
\textbf{Model} & \textbf{RMSE\,$\delta$} & \textbf{RMSE\,$\omega$} &
\textbf{Max\,$|\delta|$} & \textbf{Max\,$|\omega|$} & \textbf{Time} [s] \\ 
\midrule
HA          & 6.624e-03 & 4.948e-03 & 1.766e-02 & 1.212e-02 & 0.0180 \\
Switched    & 2.786e-02 & 1.687e-02 & 1.085e-01 & 5.938e-02 & 0.0185 \\
MLD         & 4.055e-01 & 2.078e-01 & 2.047e+00 & 1.009e+00 & 0.0056 \\
Continuous  & 5.250e-02 & 2.905e-02 & 2.316e-01 & 1.269e-01 & 0.0259 \\
\bottomrule
\end{tabular}
\label{tab:SMIB}
\end{table*}

From this example, it is observed that a continuous approximation is
sufficient when
\begin{itemize}
    \item The reset map is identity ($x^{+}=x$)
    \item Switching does not create sharp discontinuities between the
         mode-dependent vector fields
\item The effect of switching on the
dynamics is relatively weak.
\end{itemize}

In addition to these observations, ~\autoref{tab:hybrid_state_space_comparison} compares the structural composition, state-space cardinality, and growth behavior of the discussed frameworks. This quantifies how each formalism scales with the number of modes, clocks, switches, binary variables, or regions.

\begin{table*}[ht]
\centering
\caption{Structural comparison of hybrid modeling frameworks: state-space size, complexity, and growth behaviors.}
\renewcommand{\arraystretch}{1.0}
\begin{tabular}{l p{2.6cm} p{2.3cm} p{2.3cm} p{1.6cm} p{3.5cm}}
\toprule
\textbf{Model} &
\textbf{Components} &
\textbf{State Space Size} &
\textbf{Discrete Complexity} &
\textbf{Growth Rate} &
\textbf{Interpretation} \\
\midrule

HA &
\begin{tabular}[t]{@{}l@{}}%
$q\in Q$, $|Q|=M$\\
$x\in\mathbb{R}^n$
\end{tabular} &
$Q \times \mathbb{R}^n$ &
$M$ &
$\mathcal{O}(M)$ &
Each mode adds one full continuous region. \\
\midrule

Timed HA &
\begin{tabular}[t]{@{}l@{}}%
Clocks: $c\in\mathbb{R}_{\ge0}^{k}$\\
 $x\in\mathbb{R}^n$
\end{tabular} &
$Q \times \mathbb{R}^{n+k}$ &
$M$ &
$\mathcal{O}(M(n+k))$ &
Clock variables increase the continuous space; no new discrete branching. \\
\midrule

Switched &
\begin{tabular}[t]{@{}l@{}}%
 $\sigma\in\{1,\dots,S\}$\\
$x\in\mathbb{R}^n$
\end{tabular} &
$\mathcal{S} \times \mathbb{R}^n$ &
$S$ &
$\mathcal{O}(S)$ &
Linear growth with number of switch configurations. \\
\midrule

MLD &
\begin{tabular}[t]{@{}l@{}}%
 $z\in\mathbb{R}^{n_z}$\\
 $\delta\in\{0,1\}^b$
\end{tabular} &
$\mathbb{R}^{n+n_z}\times\{0,1\}^b$ &
$2^{b}$ &
$\mathcal{O}(2^b)$ &
Each binary variable doubles discrete realizations. \\
\midrule

PWA &
\begin{tabular}[t]{@{}l@{}}%
$x\in\mathbb{R}^n$\\
Regions: $\{1,\dots,R\}$
\end{tabular} &
$\displaystyle\bigcup_{i=1}^{R}\mathbb{R}^n$ &
$R$ &
$\mathcal{O}(R)$ &
One affine map per region; linear growth in $R$. \\
\midrule

Set-Oriented &
$x \in C \cup D \subseteq \mathbb{R}^n$&
$C \cup D$&
None &
$\mathcal{O}(n)$ &
Purely continuous growth; no discrete explosion. \\
\bottomrule
\end{tabular}
\label{tab:hybrid_state_space_comparison}
\end{table*}
Further, it is important to understand how these modeling choices translate into computational performance for both control and analysis synthesis, as these tasks may involve solving mixed-integer quadratic or linear optimization programs that influence scalability of the method. \autoref{tab:hybrid_mpc_reachability} provides a quantitative comparison of  formulations for MPC and reachability analysis based on HA, switched, MLD and PWA formalisms with the underlying reachable set computation methods. This is further discussed in \autoref{sec:verification}. In particular, the set-oriented method for MPC problems is typically reformulated into one of the listed hybrid representations before an optimization problem can be constructed \cite{9992597,Sanfelice2019} and, to the best of our knowledge, the approach has not yet been employed in non-analytical reachability computations with reported metrics comparable to those in \autoref{tab:hybrid_mpc_reachability}.

\begin{table*}[ht]
\centering
\scriptsize
\caption{Comparison of MPC and reachability methods based on hybrid system models.}
\label{tab:hybrid_mpc_reachability}
\renewcommand{\arraystretch}{1.2}
\begin{tabular}{p{2.7cm} p{3.2cm} p{2.8cm} p{2.9cm} p{2.5cm} }
\toprule
\textbf{Model} & \textbf{Case-Study} &
\textbf{Scalability} &
\textbf{Computation Time} &
\textbf{Complexity Growth} \\
\midrule

\multicolumn{5}{c}{\textbf{Model predictive control}} \\
\midrule

HA \cite{ha_mpc} & 126 states, 204 binaries, \newline 1536 constraints & Exponential growth in \newline mode sequences with \newline prediction horizon N.
 \newline
 &
CPLEX: 13.68 s \newline(3 iterations)\newline
Avg.\ MPC step: 3.28 s \newline
Worst-case: 23.9 s 
 & $\mathcal{O}(M^N)$\\
\midrule
Switched \cite{ChenLazar2020} & 
Linear: 2 states, 1 input, \newline 2 modes, $N=20$ \newline
Nonlinear: 6 states, 2 inputs, 2 modes, $N=40$ &
Decision size grows \newline linearly with prediction horizon and polynomially with continuous state and input dimensions.
 &
0.54--0.72 ms (linear) \newline
30--35 ms (nonlinear) \newline &
$\mathcal{O}(\mathrm{poly}(n)\,N)$ \\
\midrule

\multirow{2}{*}{MLD \cite{mld_mpc,Yaakoubi2023} } & 3 states, 1 input, 40 binaries&
Binary decision variables $\delta$ grow with prediction \newline horizon.
 &
PRESAS: 6.8--14~ms Gurobi: 8--16 ms \newline MOSEK: 126--465 ms &
$\mathcal{O}(2^{\delta N})$
 \\
\cline{2-2}\cline{4-4}

& 3 states, 6 inputs, 128 modes\newline 14 binaries,horizon $N=2$ &
 &
n/a &
 \\
\midrule

PWA \cite{Poggi2011} & 8 states $\rightarrow$ 44 regions, \newline 2-6 input horizon &
Number of regions grows with state dimension and input horizon.
\;
 &
Offline: 13.6--147.5 s \newline
Online: 26--1227 FLOPs \newline QP solver $\sim$25{,}000 FLOPs &
Offline: $\mathcal{O}(R^N)$ \newline Online: $\mathcal{O}(R)$
 \\
\midrule

\multicolumn{5}{c}{\textbf{Reachability analysis}} \\
\midrule

\multirow{2}{*}{\parbox{2.5cm}{{HA}\\ Support functions \cite{Parallel_ha}}}
 &  2 states,\newline 9, 25, 81 modes &
Problem size $\propto$ (number of support directions) $\times$ (time steps T).
&
37--217 s (9 modes) \newline
133--4835 s (25 modes) \newline
up to 56{,}590 s (81 modes) &
$\mathcal{O}(M\,T)$ \\
\cline{2-3}\cline{4-5}
Zonotopes \cite{7059096} & 2 states, \newline 6, 14 modes & Problem size $\propto$ (state\newline dimension n) $\times$ (number of zonotope generators g).
&
0.76 s (6 modes)\newline
1.95 s (14 modes)
  &
$O(n^{5})$  \\
\midrule

 \parbox{3cm}{ Switched \\Matrix-measure \cite{7040386}} & 3, 7, 15, 31, 63, \newline  127 states system&
Problem size $\propto$ (number of modes) $\times$ (time steps) &
1.52 s (3 states)\newline
3.28 s (7 states)\newline
7.60 s (15 states)\newline
17.99 s (31 states)\newline
51.41 s (63 states)\newline
107.53 s (127 states)
 &
$\mathcal{O}(S\,T)$
 \\
\midrule

\multirow{2}{*}{\parbox{2.7cm}{{MLD}\\ Polyhedral sets \cite{BemporadGiovanardiTorrisi2000}}}&  2 states, 4 regions
&
Problem size $\propto$ (number of regions) $\times$ (time steps) 
 &
39.21 s (304 QPs) \newline
Brute force: 65{,}536 QPs \ MIQP: 767 QPs &
$\mathcal{O}(2^{R}\,T)$
\\
\cline{2-2}\cline{4-5}
Zonotopes \cite{Bird2023} &  2--36 states, \newline 1--4 binary states  &
&
0.02--3269 s &
$O(n^{3})$ \\
\midrule
\parbox{3cm}{ PWA\\Ellipsoids \cite{THUAN2023101301}} &  4 and 8-state systems &
Problem size $\propto$ (number of regions) $\times$ (affine subsystem) 
 &
0.10 s (4-state) \newline 0.30 s (8-state) &
$O(n^{3})$ \\
\bottomrule
\end{tabular}
\end{table*}

This comparison exposes the scalability constraints inherent in each hybrid system model and details how their computational complexity increases. For instance, across most methods, computational effort tends to grow either with the dimension of the continuous state vector or with the number of modes and regions present in the hybrid model. In HA-based reachability methods that employ zonotopes, the cost of constructing flowpipes has been reported to exhibit a fifth-order dependence on the continuous state dimension which reflects the significant expansion of the generator sets during propagation. By contrast, systems of similar dimension exhibit a cubic dependence on the continuous state dimension in PWA and MLD frameworks. This implies that different modeling abstractions can lead to substantially different computational behaviors even when applied to systems of comparable size.

\subsubsection{Non-differentiable reset maps in model selection}
Another important modeling consideration often overlooked when selecting a hybrid framework is the ability to express reset maps in a differentiable form, which is essential when requiring access to sensitivities—such as in the computation of saltation matrices for trajectory analysis \cite{saltation}, or in hybrid extensions of Kalman filtering \cite{KONG2021109752,Odunlami2025c} and observer design. These tasks require access to the Jacobian of the reset function, or at least a consistent linearization of how state trajectories respond to discrete transitions.

Typical examples of non-differentiable resets include:
\begin{enumerate}[nosep]
\item Limiters in inverter controllers, excitation systems, and governors, which immediately clip current, voltage, or control signals to prescribed minimum or maximum values.
\item Breaker operations, which instantaneously reset voltage or current states in response to protection triggers.
\item Deadband controllers that exhibit sudden state transitions once thresholds are crossed.
\item State latching or memory elements, which introduce discrete state changes based on internal logic.
\item Anti-windup schemes, where integral states in control loops are forcibly clamped or reset to prevent instability.
\end{enumerate}

While most hybrid frameworks include some mechanism for updating the state during mode transitions, not all provide a reset map in a form that supports differentiation. In the MLD and PWA frameworks, such resets are typically encoded through logical constraints or affine switching conditions rather than expressed as explicit functions of the pre-transition state. While these models can replicate the observable effects of resets, they often do so in ways that obscure the underlying reset function or render it non-differentiable, posing challenges for tasks that rely on sensitivity analysis or gradient-based estimation.

In contrast, HA explicitly incorporates guard conditions and reset maps as part of their formal semantics, making them well-suited for capturing and analyzing non-differentiable transitions. Similarly to HA, the jump set of a set-oriented method can encode a differentiable or non-differentiable reset map. The switched system model can also accommodate such behavior if their mode transition logic is formulated to include clear state update rules. 

However, numerical stability during Jacobian propagation of non-differentiable reset map can be ensured by smoothing or regularizing the reset so that a well-defined derivative exists in a neighborhood of the jump. Hard saturation or deadband nonlinearities can be replaced by smooth approximations whose derivatives remain bounded, typically by enforcing a small positive lower bound on the slope, following domain-relaxation techniques used for non-smooth transitions \cite{6161050}. Alternatively, directional derivatives within a narrow guard band around the reset surface or Jacobian-free finite-difference evaluations \cite{KnollKeyes2004} may be used. Similarly, some estimation methods avoid differentiating the reset altogether by modeling its effect as heavy-tailed process noise \cite{Zhang2020HybridEstimation}.

\subsection{Comparative study of HDS and purely continuous for modeling and estimation}
In this section, we compare the performance of purely continuous and HDS formulations for power systems state estimation.

\subsubsection{Grid-connected inverter under mode switching}
Consider a grid-connected inverter whose underlying dynamics switch between two fundamentally different operating modes: grid-following (GFL) and grid-forming (GFM) \cite{Odunlami2025c,GFL_GFM}. The goal is to estimate the inverter's internal states from noisy measurements and to determine the estimation accuracy when the assumed model is either (i) hybrid with explicit mode switching and reset logic, or (ii) a continuous model that smooths over switching events. Mode transitions are triggered by the measured grid voltage magnitude $\lvert V_{\text{grid}} \rvert$ crossing predefined thresholds. When $\lvert V_{\text{grid}} \rvert$ falls below a low-voltage threshold, the inverter switches from GFL to GFM mode, as in a low-voltage ride-through activation. When $\lvert V_{\text{grid}} \rvert$ rises above a higher threshold, the system transitions back from GFM to GFL mode, representing post-fault grid reconnection. A hysteresis band between low and high voltage thresholds is used to prevent chattering.

GFL inverters are a vital component of modern power systems, allowing the connection of renewable energy resources to the grid. Understanding their dynamic behavior is crucial for enhancing performance and ensuring stability, particularly as power systems evolve to accommodate increasing levels of renewable generation. The GFL inverters operate in a current-controlled mode, regulating the injected active and reactive power while synchronized to the grid voltage through a phase-locked loop. Here, the inverter is modeled in the synchronous $dq$ frame with the $d$-axis aligned to the grid voltage. The dynamics of the output currents are given by:
\begin{align}
\frac{d}{dt}{i}_d &= \frac{v_d - R_{\mathrm{pu}} i_d + \omega L_{\mathrm{pu}} i_q}{L_{\mathrm{pu}}} \label{eq:gfl_id}\\
\frac{d}{dt}{i}_q &= \frac{v_q - R_{\mathrm{pu}} i_q - \omega L_{\mathrm{pu}} i_d}{L_{\mathrm{pu}}} \label{eq:gfl_iq}
\end{align}
where $i_d,i_q$ are the inverter output currents, $v_d,v_q$ are the inverter terminal voltages in the $dq$ frame, $R_{\mathrm{pu}}$ and $L_{\mathrm{pu}}$ are the per-unit filter resistance and inductance, and $\omega$ is the grid angular frequency.

Conversely, GFM inverters operate under voltage control, providing and regulating local voltage and frequency references. They are capable of maintaining stable operation in both grid-connected and islanded conditions, providing system support during disturbances. In GFM mode, the inverter controls the terminal voltage and frequency using a droop-based strategy. The dynamics of the current and voltage are governed by:
\begin{align}
\frac{d}{dt}{v}_d &= \frac{\omega L_{\mathrm{pu}} i_q - R_{\mathrm{pu}} i_d}{L_{\mathrm{pu}}} \label{eq:gfm_vd}\\
\frac{d}{dt}{v}_q &= \frac{-\omega L_{\mathrm{pu}} i_d - R_{\mathrm{pu}} i_q}{L_{\mathrm{pu}}} \label{eq:gfm_vq}\\
i_d &= \frac{V_{\mathrm{ref}} - v_d}{R_{\mathrm{pu}}} \label{eq:gfm_id}\\
i_q &= -\frac{v_q}{R_{\mathrm{pu}}}\label{eq:gfm_iq}
\end{align}
where $V_{\mathrm{ref}}$ is the $d$-axis voltage reference in GFM mode.

The inverter with mode switching is represented by an HA, which explicitly models mode switching via guard conditions, applies state resets when transitioning from GFL to GFM operation, and propagates covariance across the switching surface using the saltation matrix. In contrast, the continuous formulation employs a smooth sigmoid function of the grid voltage magnitude to interpolate between GFL and GFM dynamics.
The reset map in the HA clamps the current states $(i_d, i_q)$ to the GFM current limit when transitioning from GFL to GFM, leaving the voltage states $(v_d, v_q)$ unchanged, and no reset is applied when switching from GFM to GFL.

The continuous and hybrid models are embedded within an extended Kalman filter and tested under identical conditions. Following \cite{7742899}, the extended Kalman filter is implemented as a two-step recursive algorithm consisting of a prediction step and a correction step. In the prediction step, the nonlinear inverter dynamics are propagated forward in time using a fourth-order Runge-Kutta integration scheme. The associated state transition Jacobian is obtained numerically and used to update the error covariance matrix. The correction step then incorporates noisy measurements to refine the state estimate. The EKF recursion in both formulations follows a similar structure and can be expressed as

\begin{align}
x_{k|k-1} &= f(x_{k-1|k-1}, u_{k-1}) \\
P_{k|k-1} &= F_{k-1} P_{k-1|k-1} F_{k-1}^\top + Q  \\
K_k &= P_{k|k-1} H^\top \big( H P_{k|k-1} H^\top + R_y \big)^{-1} \\
x_{k|k} &= x_{k|k-1} + K_k \big( z_k - H x_{k|k-1} \big) \\
P_{k|k} &= \big( I - K_k H \big) P_{k|k-1}
\end{align}

\noindent where the subscript $k$ denotes the current discrete time step and $k-1$ the previous step, while $x_{k|k-1}$ and $x_{k|k}$ refer to the predicted and corrected state estimates, respectively. $P$ is the error covariance matrix, $F$ is the Jacobian of the nonlinear state transition model, $K$ is the Kalman gain, and $H$ is the measurement matrix. 

The distinction between the two formulations arises only in the prediction step of the EKF. In the hybrid model, the process dynamics switch explicitly between GFL and GFM equations depending on the grid voltage. At a mode transition, the predicted state is updated by the reset map $\mathcal{R}(\cdot)$, and the error covariance is propagated through the \textit{saltation matrix}~$\Xi$ \cite{KONG2021109752},  
\begin{equation}
P_{k|k-1} \;\leftarrow\; \Xi(x_{k|k-1})\, P_{k|k-1}\, \Xi(x_{k|k-1})^\top
\end{equation}
where $\Xi$ extends the Jacobian of the reset map $D_x\mathcal{R}$ by incorporating terms that capture the discontinuity of the vector fields across the guard surface, ensuring a first-order consistent mapping of uncertainty at the switching instant.  

In the continuous model, the prediction step employs a smooth approximation between GFL and GFM dynamics such that:  
\begin{equation}
f^{\mathrm{cont}}(x) \;=\; \sigma(V_{\mathrm{grid}})\, f_{\mathrm{GFL}}(x) 
\;+\; \big(1-\sigma(V_{\mathrm{grid}})\big)\, f_{\mathrm{GFM}}(x)
\end{equation}
with $\sigma(V_{\text{grid}})=\dfrac{1}{1+\exp\!\big[-\Gamma \,(V_{\text{grid}}-V_{\text{th}})\big]}$

\noindent where $\Gamma$ is a gain parameter that controls the sharpness of the transition and $V_{\mathrm{th}}$ denotes the switching threshold voltage. 

\begin{figure*}[ht]
\centering
\includegraphics[width=0.8\linewidth]{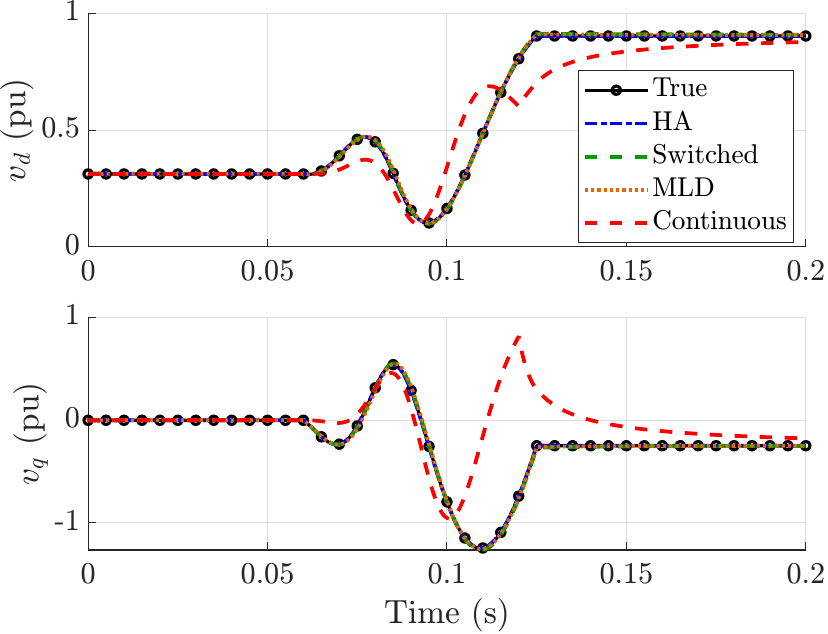}
\caption{Variation of voltage in d- and q-axes for hybrid and continuous models.}
\label{fig:estimated_voltage}
\end{figure*}

\begin{figure*}[!t]
\centering
\includegraphics[width=0.78\linewidth]{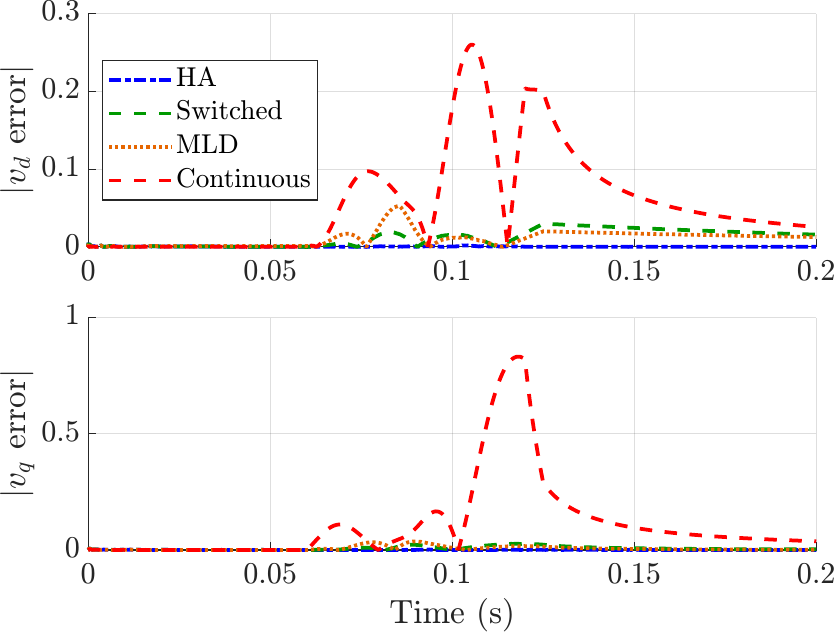}
\caption{Performance comparison of absolute voltage estimation errors in the d- and q-axes for hybrid and continuous models.}
\label{fig:voltage_errors}
\end{figure*}

\begin{figure*}[!t]
\centering
\includegraphics[width=0.82\linewidth]{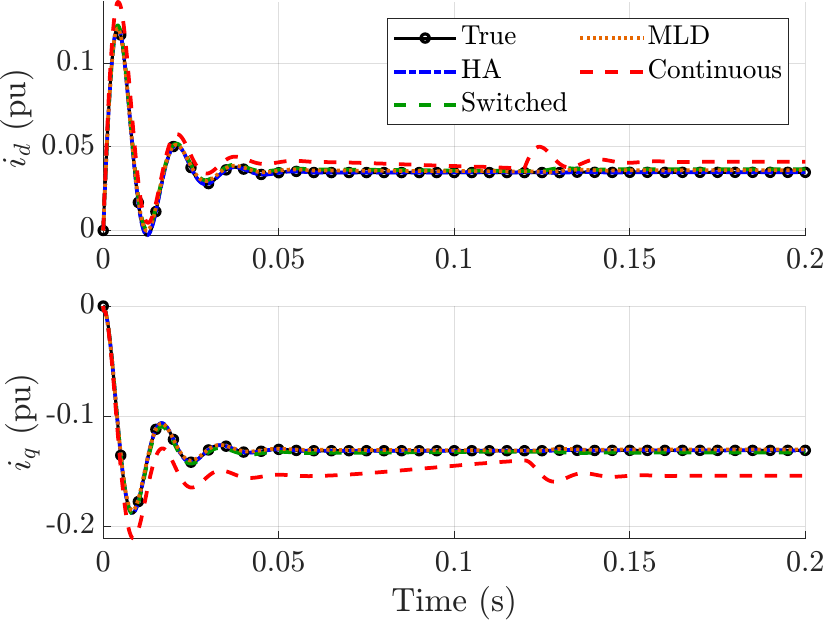}
\caption{Variation of current in d- and q-axes for hybrid and continuous models.}
\label{fig:estimated_currents}
\end{figure*}

\begin{figure*}[!t]
\centering
\includegraphics[width=0.85\linewidth]{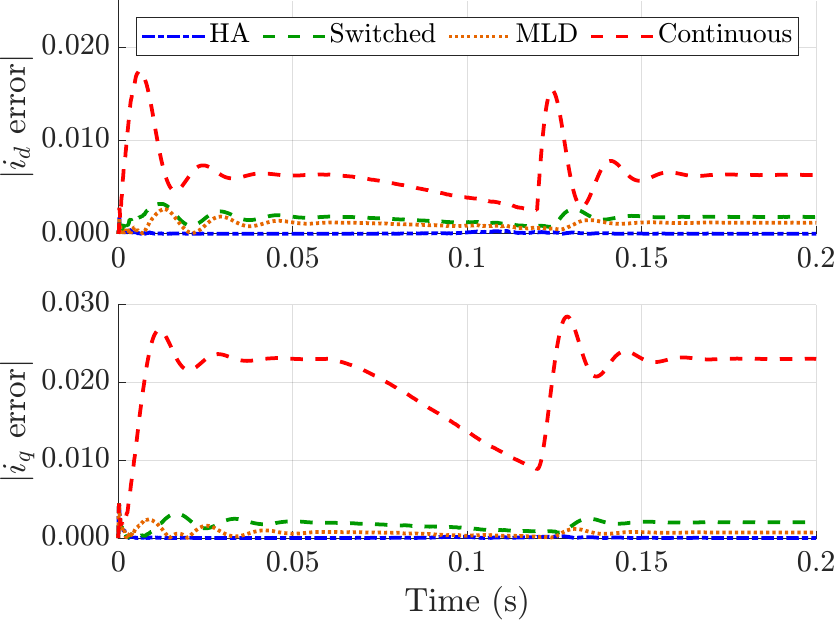}
\caption{Performance comparison of absolute current estimation errors in the d- and q-axes for hybrid and continuous models.}
\label{fig:current_errors}
\end{figure*}

 The estimation results are shown in Figs. \ref{fig:estimated_voltage}--\ref{fig:current_errors}.

It is evident from Figs. \ref{fig:estimated_voltage}--\ref{fig:voltage_errors} that the continuous model exhibits large transient deviations during switching events due to its inability to capture the instantaneous state reset at mode transitions. In contrast, the hybrid models align closely with the true values for both $d$- and $q$-axis voltages. \cite{DeCarlo2000} provides further evidence that such a mismatch in the continuous model can lead to incorrect conclusions about the system’s transient stability and reduce the accuracy of model-based monitoring or control.

\autoref{modeswitching} shows the performance of each modeling framework quantitatively  using the root-mean-square error (RMSE) between estimated and true values over the full simulation and in a near-switching time window, providing an objective basis for assessing how explicit mode-switch handling affects estimation outcomes, especially during localized, high-transient intervals. 
To complement the RMSE analysis in \autoref{modeswitching}, we further evaluate the four estimators using statistical fit measures and computational efficiency; see \autoref{tab:statistics}.  The coefficient of determination $R^{2}$, which measures the
fraction of the variance of the true trajectory explained by the estimator is computed for the four modeling frameworks. HA framework
provides the strongest fit under both metrics: its $R^{2}$ values are close to
unity, whereas the switched and MLD models exhibit only minor degradation  below $1\%$.
For the continuous formulation, its $R^{2}$ scores in the
direct–axis variables reduces by almost $25\%$, and those in the quadrature–axis variables become negative.
\begin{table*}[ht]
\centering
\scriptsize
\caption{RMSE comparison for extended Kalman filter estimation using three hybrid-system models and a continuous model, evaluated over near-switch and overall windows.}
\label{tab:rmse_comparison}
\begin{tabular}{lcccccccc}
\hline
& \multicolumn{4}{c}{\textbf{Near-switch RMSE}} 
& \multicolumn{4}{c}{\textbf{Overall RMSE}} \\
\cline{2-5} \cline{6-9}
\textbf{Model} 
& $i_d$ & $i_q$ & $v_d$ & $v_q$ 
& $i_d$ & $i_q$ & $v_d$ & $v_q$ \\
\hline
HA
& $2.76\times10^{-5}$ & $4.18\times10^{-5}$ & $5.42\times10^{-4}$ & $6.86\times10^{-4}$ 
& $3.06\times10^{-4}$ & $2.89\times10^{-4}$ & $7.96\times10^{-4}$ & $8.56\times10^{-4}$ \\

Switched 
& $1.65\times10^{-3}$ & $1.70\times10^{-3}$ & $1.58\times10^{-2}$ & $1.28\times10^{-2}$ 
& $1.75\times10^{-3}$ & $1.82\times10^{-3}$ & $1.54\times10^{-2}$ & $9.42\times10^{-3}$ \\

MLD
& $1.04\times10^{-3}$ & $6.97\times10^{-4}$ & $1.20\times10^{-2}$ & $1.14\times10^{-2}$ 
& $1.18\times10^{-3}$ & $8.53\times10^{-4}$ & $1.46\times10^{-2}$ & $1.05\times10^{-2}$ \\

Continuous
& $6.65\times10^{-3}$ & $2.07\times10^{-2}$ & $3.63\times10^{-1}$ & $6.36\times10^{-1}$ 
& $6.83\times10^{-3}$ & $2.10\times10^{-2}$ & $2.79\times10^{-1}$ & $4.14\times10^{-1}$ \\
\hline
\end{tabular}
\label{modeswitching}
\end{table*}
\begin{table*}[ht]
\centering
\scriptsize
\caption{Performance comparison of HA, Switched, MLD, and Continuous underlying models based on $R^2$ score, mean normalized innovation squared (NIS), and total computation time.}
\label{tab:ekf_stats}
\begin{tabular}{lcccccccccc}
\hline
\textbf{Model} 
& $R^2(i_d)$ & $R^2(i_q)$ & $R^2(v_d)$ & $R^2(v_q)$ 
& Mean NIS 
& Total time (s) \\
\hline
HA        & 0.9995 & 0.9997 & 1.0000 & 1.0000 & 4.016      & 0.0983 \\
Switched  & 0.9836 & 0.9880 & 0.9997 & 0.9993 & 7.763      & 0.0924 \\
MLD       & 0.9926 & 0.9974 & 0.9998 & 0.9991 & 7.478      & 0.1190 \\
Continuous& 0.7503 & -0.6053 & 0.9112 & -0.3587 & 40.266   & 0.0892 \\
\hline
\end{tabular}
\label{tab:statistics}
\end{table*}

The normalized innovation squared, given by
\begin{align}
\mathrm{NIS}_{k}
&=
\bigl(z_{k} - H \hat{x}_{k|k-1}\bigr)^{\top}
\bigl(H P_{k|k-1} H^{\top} + R_{y}\bigr)^{-1} \notag \\
&\quad \times
\bigl(z_{k} - H \hat{x}_{k|k-1}\bigr)
\end{align}
 tests the match between the predicted covariance and the observed innovations. For a correctly tuned and structurally adequate EKF, the NIS should behave as a 
$\chi^{2}_{n_{y}}$ random variable with $n_{y}$ degrees of freedom and its empirical mean should be close to $n_{y}$ \cite{Chen2018WeakInTheNEES}. HA yields an average NIS near the theoretical expectation and remains within the
95\% $\chi^{2}_{4}$ bounds. The switched and MLD models show moderate inflation,  with average NIS about $80$–$100\%$ above the nominal value.
In sharp contrast, the larger RMS of the continuous approach is consistent with  its average NIS, which exceeds the theoretical
value by more than two orders of magnitude.

Further, computation times indicate that the HA, switched, and continuous methods incur nearly
identical costs (under $0.1\ \mathrm{s}$), whereas the MLD model requires around $20$–$30\%$ additional
runtime due to the logical constraints imposed by the saturation rules.

\section{Safety verification and reachability analysis} \label{sec:verification}
The notion of hybrid system reachability has been gaining increasing attention in power systems, particularly as analytical methods may struggle to address the complex, nonlinear, and hybrid dynamics inherent in modern power grids \cite{7963334,7273785}. Considering the safety-critical nature of power systems, it is essential to ensure that a developed model not only comprehensively represents the system's dynamics but also adheres to safety and operational specifications. This process is often referred to as verification.

\textbf{Definition 3.} (Safety verification):
Given a hybrid system \( \mathcal{H} \) and a set of unsafe states \( U \subseteq \mathbb{R}^n \), the \textit{safety verification problem} is to determine whether there exists an initial state of \( \mathcal{H} \), \( x_0 \in X_0 \subseteq \mathbb{R}^n \) and a time \( t \geq 0 \) where the state \( x(t; x_0) \) of the system is in \( U \).

Given the limitations (e.g., model checking and theorem proving) of formal verification methods developed for discrete operations, these methods are not directly applicable for verifying power systems modeled as hybrid systems. This is because they involve exhaustive search of all states that can be reached from the initial state, and hybrid systems also consist of continuous dynamics with infinite and uncountable states. This makes exhaustive state exploration impractical and often impossible. Reachability analysis, on the other hand, focuses on computing the set of reachable states over time, often using over-approximations and efficient numerical techniques. 

\textbf{Definition 4.} (Reachable set): 
Consider a dynamical system represented by an implicit DAE:
\begin{equation}
0 = F (\dot{x}(t), x(t), u(t))
\label{eq:dae}
\end{equation}
where \( x(t) \) represents the system states and \( u(t) \) denotes the system inputs. The reachable set over the time interval \( t \in [0, t_f] \), starting from an initial set of states \( R(0) \) and subject to input uncertainties \( U \), is given by:
\begin{equation}
R(t_f) =
\left\{ x(t)\, \Bigg|\, 
\begin{aligned}
&F(\dot{x}(t), x(t), u(t)) = 0, \ t \in [0, t_f], \\
&(\dot{x}(0), x(0), u(t)) \in R(0) \times U
\end{aligned}\right.
\end{equation}
The notion of reachable sets has been extensively used in purely continuous-time systems as a safety verification method and extended to HDS. It determines whether the system's state can remain within a specified safe region over a given time horizon. Fundamentally, it aims to categorize states into two groups: those that can be reached from the initial conditions and those that cannot \cite{Althoff2014}. \autoref{tab:reachable sets in hybrid models of power systems} organizes the few methods that have been used to compute reachable sets for hybrid system models of power systems. Given that the safety verification problem is posed as a property of reachable sets \cite{Tomlin2003}, it becomes feasible to design power systems as hybrid models using a reachability-based approach. This methodology ensures that they satisfy reachability specifications. This is further discussed in literature reviews of power system applications in \autoref{subsec:verification_methods}. The reader interested in verification of general hybrid systems is referred to \cite{alur2009formal,tabuada2009verification}. Benchmarks for verification of power systems modeled as hybrid systems are also proposed in \cite{ARCH22:Benchmarks_Formal_Verification_Power,larsson2006benchmark}.
\begin{table*}[ht]
\caption{\raggedright Methods used for computing reachable sets in hybrid models of power systems}
\label{tab:reachable sets in hybrid models of power systems}
\centering
\resizebox{\textwidth}{!}{
\begin{tabular}{llll}
\Xhline{0.5pt}
\textbf{Method} & \textbf{Applications in power systems} & \textbf{Hybrid systems modeling framework} & \textbf{Ref.} \\ \Xhline{0.8pt}

Level set analysis & 1. Transient stability analysis in a transmission system & HA & \cite{Susuki2012} \\
 & 2. Fault release control verification & HA & \cite{Susuki2012} \\ 
 & 3. Voltage safety assessment & DAE with discrete state resets & \cite{10.1007/978-3-540-78929-1_51} \\ 
 & 4. Prediction of transient instability & HA & \cite{Susuki2006} \\ 
\addlinespace[1em]

Time-reverse trajectories & Prediction of voltage instability & HA & \cite{susuki2007predicting} \\ 
\addlinespace[1em]
Lagrangian-numerical integration & Fault release control verification & HA & \cite{Susuki2008} \\ 
\addlinespace[1em]

Rapidly-exploring random tree (RRT) & Dynamic security analysis (transient stability and frequency response) & HA & \cite{QiangWU2016} \\ 
\addlinespace[1em]
Sampling based-RRT & Dynamic security analysis of cyber-physical power system & HA & \cite{QWu2018} \\ 
\addlinespace[1em]

Continuation-grazing analysis & Vulnerability assessment under uncertainty & DAE with switching and state resets & \cite{inproceedings} \\ 
\Xhline{0.8pt}
\end{tabular}}
\end{table*}

\begin{table*}[!t]
\scriptsize
\centering
\caption{Applications of HDS frameworks in power systems}
\label{tab:application of HDS research in power system}
\renewcommand{\arraystretch}{1.2}
\begin{adjustbox}{width=\linewidth}
\begin{tabular}{@{}p{2.5cm} >{\centering\arraybackslash}m{4.5cm} >{\centering\arraybackslash}m{3.9cm} p{4.2cm}@{}}
\toprule
\textbf{Area} & \textbf{Specific application} & \textbf{Model/Method} & \textbf{Case study} \\ 
\midrule
\multirow{2}{*}{\makecell[c]{Model\\formulation}} & Large disturbance modeling \cite{hiskens2000b,hiskens2001,Hiskens2004} & DAD \cite{hiskens2000b}, DSAR \cite{Hiskens2004}, HA(DAIS) \cite{Hiskens2004} & AVR \& on-load tap changers \cite{hiskens2000b,hiskens2001,Hiskens2004}\\ 
\cline{2-4}
& Power grid side converter \cite{fan2008hybrid} & HA & Double-fed wind power generator \\
\cline{2-4}
& Electric power transmission \cite{fourlas2004,fourlas2005modeling} & HIOA & 4-bus, 6 transmission line system \\
\cline{2-4}
& Performance evaluation of renewable energy resources \cite{guerin2011hybrid} & DAE+switching logic & Solar photovoltaics + wind turbine + battery energy storage system \\
\cline{2-4}
& MVDC shipboard power system \cite{babaei2015development} & HA & Shipboard power system\\
\cline{2-4}
& Marine power plant control \cite{miyazaki2016hybrid} & HDS framework in (1) & Marine power plant\\
\cline{2-4}
& Energy storage\&load \cite{silva2018hybrid, hybridPetrinet} & DHA, PWA, MLD \cite{silva2018hybrid}, HPN \cite{hybridPetrinet} & Pilot microgrid \cite{silva2018hybrid}, 1MW renewable plant with storage \cite{hybridPetrinet} \\
\cline{2-4}
& Virtual power plant modeling \cite{javaid2019modelling} & HA & VPP at Newcastle University\\
\cline{2-4}
& Cyber-physical distribution grid modeling \cite{9632321} & HA & Low-voltage distribution grid with solar photovoltaics\\
\cline{2-4}
& Controlled line switching \cite{ZahnHagenmeyer} & FHA & 3-bus AC microgrid \\
\midrule
\multirow{2}{*}{\makecell[l]{State/parameter \\ estimation}} & Parameter identifiability \cite{hiskens2000Identifiability} & HA & on-load tap changers\\ 
\cline{2-4}
& Moving horizon estimation - fault detection \cite{thomas2003moving} & MLD & Steam generator \\
\cline{2-4}
& System identification \cite{windpower} & PWA & 500kW hydraulic wind power system \\
\midrule
\multirow{5}{*}{\makecell[l]{Stability \\ and \\ control}} &
Transmission system transient stability \cite{susuki2005application,Koo2009} & HA/Hybrid reachability analysis & SMIB \cite{susuki2005application}, SMIB \& DMIB \cite{Koo2009} \\ 
\cline{2-4}
& Voltage instability prediction \cite{susuki2007predicting} & HA/Hybrid reachability analysis & Single machine load bus\\ 
\cline{2-4}
& Voltage control \cite{beccuti2005hybrid,moradzadeh2010hybrid,nguyen2015hybrid,Geyer2003,Negenborn2007} & MLD/MPC \cite{beccuti2005hybrid,Geyer2003,Negenborn2007}, HA \cite{nguyen2015hybrid,moradzadeh2010hybrid} & 12-bus system \cite{beccuti2005hybrid,moradzadeh2010hybrid}, SVC \cite{nguyen2015hybrid}, ABB three-area system \cite{Geyer2003}, IEEE 9-bus system \cite{Negenborn2007}\\ 
\cline{2-4}
& Controller design \cite{1713273} & HPN + HA & 1300~MW pumped-storage power station\\ 
\cline{2-4}
& Global control \& state management \cite{4620733} & HA & 300kw-grid connected WTG\\ 
\cline{2-4}
& Power management system control \cite{kafetzis2020energy,reddy2023decentralized, 4919363} & HA \cite{kafetzis2020energy}, HDS framework in (1) \cite{reddy2023decentralized}, MLD \cite{4919363} & 100kW off-grid microgrid \cite{kafetzis2020energy}, ship power system \cite{reddy2023decentralized}, 50kW residential smart grid \\ 
\cline{2-4}
& Energy control \cite{HybridMPC,10.1002/asjc.1967} & MLD/MPC & DC microgrid with solar photovoltaics + battery \cite{HybridMPC}, large scale solar field \cite{10.1002/asjc.1967} \\
\cline{2-4}
& Voltage stability evaluation and control \cite{https://doi.org/10.1155/2023/5037957} & HA & WSCC 3-machine, 9 bus system. 10-machine, 39-bus New England power system\\ 
\midrule
\multirow{2}{*}{\makecell[c]{Anomaly \\ detection \\and location}} & Fault detection and isolation \cite{fourlas2001framework,fourlas2002diagnosability} & HIOA & 100 km two-line transmission system\\ 
\cline{2-4}
& Fault diagnosis \cite{Fourlas2005diagnoser,wang2005hybrid,8535711} & HIOA \cite{Fourlas2005diagnoser}, CGHDS \cite{wang2005hybrid}, Stochastic HA \cite{8535711} & 4-bus - system with circuit breakers and inverse time relays \cite{Fourlas2005diagnoser}, WSCC-9 bus system \cite{wang2005hybrid}, lithium-ion battery system \cite{8535711} \\ 
\cline{2-4}
 & Cyber-security (false data injection attack) \cite{FDIAattack} & HA/reachability analysis & DC microgrid\\ 

\midrule

\multirow{2}{*}{\makecell[l]{Dynamical \\ analysis}} & Cascading failure analysis \cite{Susuki2009,YangEvent} & HA/reachability analysis \cite{Susuki2009}. Event-triggered HA \cite{YangEvent} & Six-machine power system model \cite{Susuki2009}. 9-bus, 39-bus, 200-bus system \& 2000-bus Texas network \cite{YangEvent} \\ 
\cline{2-4}
& Trajectory sensitivity analysis \cite{hiskens2000trajectory,8684267} & DAD \cite{hiskens2000trajectory}, DAIS \cite{8684267} & AVR + on-load tap changers \cite{hiskens2000trajectory}, SMIB \cite{8684267} \\ 
\midrule
\multirow{1}{*}{\makecell[l]{Optimization}} & Optimal power management \cite{kwatnyship,5721189} & HA/MIDP \cite{kwatnyship}, MLD/MIQP \cite{5721189}& Ship power system \cite{kwatnyship}, solar photovoltaic system \cite{5721189} \\ 
\cline{2-4}
& Power system stabilizer optimization \cite{baek2008power} & DAIS-FFNN & SMIB, MMPS \\ 
\cline{2-4}
& Renewable hybrid power systems \cite{wu2018} & Switched system & 200kW microgrid with wind, solar and battery storage \\
\cline{2-4}
& Co-generation plant scheduling \cite{Ferrari-Trecate2004} & MLD/MPC & Gas turbine + heat recovery steam generator + auxiliary boilers \\
\midrule
\multirow{7}{*}{\makecell[c]{Verification}} 
& Fault release control verification \cite{Susuki2008} & HA/forward reachability analysis & DMIB \\ 
\cline{2-4}
& Dynamic security assessment \cite{inproceedings} & DAE + switching and state resets & IEEE 14- bus system \\ 
\cline{2-4}
& Smart grid safety verification \cite{Shaid2014} & Stochastic HA & Smart grid \\ 
\cline{2-4}
& Protection schemes verification \cite{6975219} & HA/temporal logic & IEEE-9 bus system \\ 
\cline{2-4}
& Flexible load control in active distribution networks \cite{8245721} & HA & Combined cooling, heat and power system \\ 
\bottomrule
\end{tabular}
\end{adjustbox}
\end{table*}

\section{Applications in power systems} \label{sec:applications}
\begin{figure*}[ht!]
\centering
\includegraphics[width=1\linewidth]{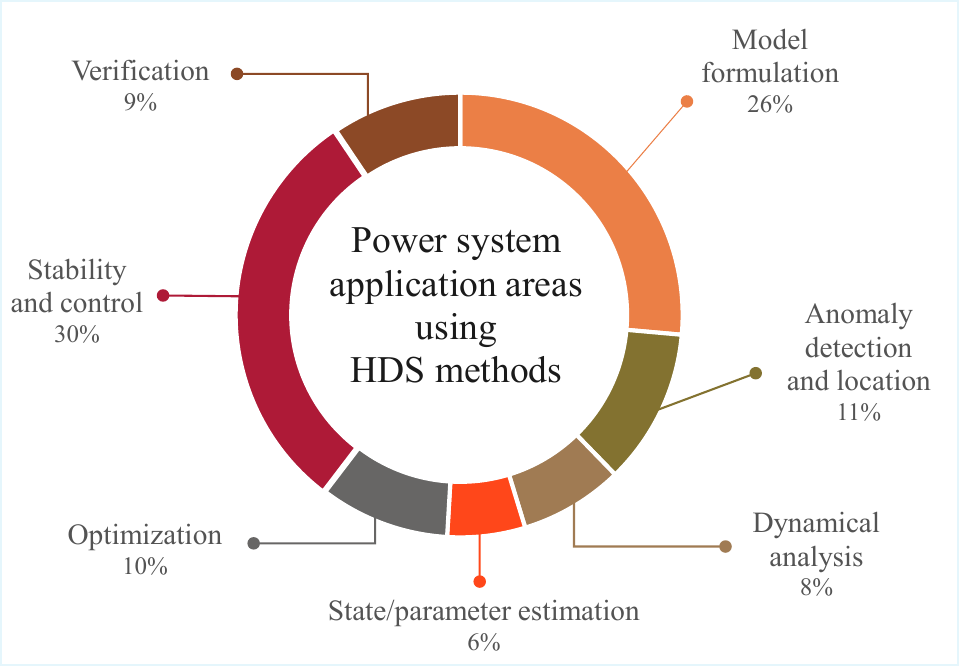}
\caption{Distribution of articles across different application areas in power systems as summarized in \autoref{tab:application of HDS research in power system}.}
\label{fig:articlebyarea}
\end{figure*}

Research in HDS theory is typically grouped into four broad categories: modeling, analysis, control, and design \cite{branicky1996class}. In this work, we highlight power systems applications that go beyond general problem formulations. \autoref{tab:application of HDS research in power system} categorizes these applications by area, specifying the use case, the models or methods employed, and the corresponding case studies. The range of applications includes model development, state and parameter estimation, stability analysis and control, anomaly detection and localization, dynamic behavior analysis, optimization, and verification. 
 The distribution of surveyed articles is provided in \autoref{fig:articlebyarea}. This section provides a careful review of the details and rationale underlying relevant literature, including foundational works and prominent applications. It also highlights potential
 research directions based on the drawbacks and characteristics of these applications.
 \subsection{Model development}
\label{subsec:model_development}
Positive sequence (phasor domain) models are the most widely adopted for power system dynamics. These models take the form of DAEs and face several challenges. Discrete events cause abrupt changes in the algebraic constraints and may alter the DAE index, which can lead to numerical instability. Estimation tasks are sensitive to the choice of measured variables and errors in the initialization of algebraic states can propagate throughout the coupled system.

\noindent Hybrid extensions have therefore been proposed to embed logic-driven transitions directly within the dynamic structure. The differential-algebraic-discrete (DAD) formulation \cite{hiskens2000b} represents an early step in this direction, integrating continuous, algebraic, and  discrete variables to describe switching actions such as tap changes and protective relay operations within a unified framework. 
 Subsequent refinements emphasize more systematic handling of switching and state updates. The differential, switched algebraic and state-reset (DSAR) formulation \cite{hiskens2001} extends the DAD approach by adding switched algebraic relations and explicit state resets, enabling large-disturbance simulations to be treated within a single differential–algebraic structure. The differential-algebraic impulsive-switched (DAIS) formulation \cite{Hiskens2004} further generalizes this approach by admitting impulsive actions and providing a clearer representation of discontinuous transitions such as tap-step changes.

 While the assumption of fixed state dimension simplifies implementation, each event still requires reinitialization and recomputation of interactions, which increases computational effort in large systems. Hybrid automata-based approaches provide a formal structure for representing interactions among subsystems. The hybrid input/output automata (HIOA) framework \cite{fourlas2004} organizes variables into input, internal, and output sets and supports the composition of multiple automata,
\begin{equation}
H_1 \times H_2 = (U_1 \cup U_2, X_1 \cup X_2, Y_1 \cup Y_2, \dots)
\label{eq:COMPOSITIONHA}
\end{equation}
\noindent thereby enabling larger models to be constructed from subsystem descriptions and facilitating modular representations of load dynamics, relay behavior, and system operating states \cite{Liacco1978, Fink1978}. Related hybrid formulations have also been proposed in areas such as converter switching, microgrid coordination, and voltage regulation \cite{fan2008hybrid, silva2018hybrid, 9632321}.

More recent developments add control-theoretic structure. The flat hybrid automata (FHA) framework \cite{ZahnHagenmeyer} embeds differential flatness into each discrete mode so that state and input trajectories can be parameterized by flat outputs that simplify planning and control within each mode, whle guards, resets, and mode transitions remain consistent with standard HA structure. 
\begin{figure*}[!t]
\centering
\includegraphics[width=0.7\linewidth]{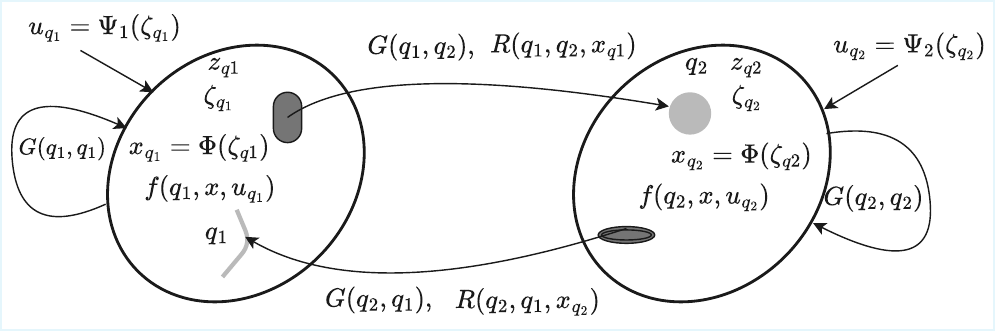}
\caption{Elements of a FHA with two discrete modes $q_1$ and $q_2$, including self-transitions and mode-to-mode transitions.}
\label{fig:FHA}
\end{figure*}

\textbf{Definition 5.} (Differential flatness):  
A system is differentially flat if there exists a flat output \( z \) such that the states \( x \) and inputs \( u \) can be expressed as functions of \( z \) and its derivatives:
\( x = \Phi(z, \dot{z}, \ddot{z}, \dots, z^{(\gamma)}) \) and
\( u = \Psi(z, \dot{z}, \ddot{z}, \dots, z^{(\gamma+1)}) \),
where \( \gamma \) is the highest derivative required. The system can be transformed into a Brunovský form
\( \dot{\zeta} = g(\zeta, \nu) \),
where \( \zeta = (\dot{z}, \ddot{z}, \dots, z^{(\gamma)}) \) represents the flat state.

\subsubsection{Potential research directions} \label{subsubsec:model_summary}
The reviewed literature shows clear progress in hybrid modeling for renewable energy systems and other power system applications, reflecting a shift from continuous-time DAEs to richer hybrid formulations. While these developments enhance the representation of switching and multi-modal behavior, many remain case-specific, and the lack of standardization complicates broader adoption. Several challenges remain open. Hybrid simulations are computationally demanding, particularly when frequent state resets, event timing, and continuous–discrete interactions must be resolved accurately. Large-scale networks, such as power systems, are especially sensitive to exponential growth in the number of variables, emphasizing the need for scalable formulations and modular designs that incorporate diverse components without excessive computational cost.

\subsection{State and parameter estimation}
\label{subsec:state_estimation}

Given the inherent variability and uncertainty in renewable energy sources, meaningful modeling of HDS requires precise identification of state and model parameters. The study in \cite{hiskens2000Identifiability} is the first to examine this problem in the hybrid power-system setting, providing a theoretical framework for analyzing when parameters can be uniquely recovered. The paper characterizes conditions under which estimation may fail and shows that recoverability depends on the linear independence of the associated trajectory sensitivities

Hybrid estimation methods have since evolved along two main directions. The first involves optimization-based estimators formulated on hybrid models. An early example is the moving-horizon approach in \cite{thomas2003moving}, which employs the MLD framework to model a steam generator, thus enabling sensor fault detection. The estimator enforces consistency with both continuous dynamics and discrete sensor logic while minimizing output mismatch over a finite horizon. To manage computational burden, the authors divide sensors into subsets, solve each subset as a mixed-integer quadratic program, and detect faults using residual thresholds. This demonstrates the feasibility of real-time hybrid-model-based estimation, while highlighting trade-offs between computational effort, estimator horizon, and observability.

The second direction concerns identification of hybrid or piecewise models for systems exhibiting strong nonlinearities. A representative example is the hydraulic wind power transfer system in \cite{windpower}, where actuator hysteresis and operating-region variability render a single linear model inadequate. A PWA identification scheme is used to determine operating regions and parameters. The result shows improved state estimation with over 91\% consistency with experimental measurements.

\subsubsection{Potential research directions}
\label{subsubsec:state_summary}

The identification problem in hybrid systems involves determining unknown states or parameters and inferring unobserved switching times and hidden operating modes. The analysis in \cite{hiskens2000Identifiability} established conditions for local identifiability based on trajectory sensitivities, but did not address global guarantees, strong nonlinearities, or sparse measurements. Subsequent work \cite{Paoletti2007, MORADVANDI202395} covers parts of this gap, but applications in power systems remain limited and still largely centered on PWA formulations. The availability of diverse
identification methods is particularly important in power systems, where it is impractical to model every
aspect purely from first principles and many industrial applications rely on output-only models to understand the large-scale system dynamics. In such cases, data-driven approaches provide a valuable means of capturing system
behaviour. Another direction lies in optimizing the MLD formalism and the moving horizon estimation framework to
address the increasing scale and complexity of modern power systems Improvements to MLD and moving-horizon formulations are required for large-scale networks. As noted in \cite{thomas2003moving}, longer horizons enhance estimation accuracy but impose prohibitive computational cost, and sensor-wise decomposition can undermine observability.

\subsection{Stability and control} \label{subsec:stability_control}

A fundamental requirement in power systems is ensuring stability, particularly in response to faults or other disturbances, across different operating points. Traditional single–mode tools have difficulty representing how discrete events can abruptly redirect system trajectories, motivating the use of HDS models for stability assessment and control design.

A recurring direction in the literature is the use of hybrid reachability analysis to examine post-fault behavior and evolving stability regions. HA formulations \cite{susuki2005application,Koo2009} describe both generator dynamics and discrete interventions, showing fault clearing and line switching increases the set of admissible operating points. 

Another prominent direction involves hybrid predictive control for voltage stability. MLD-based MPC schemes \cite{Geyer2003,Negenborn2007} couple continuous voltage evolution with event-driven changes in network configuration within one optimization framework. These works emphasize anticipating collapse trajectories and show that coordinated hybrid control can prevent voltage instability in situations where open-loop operation would otherwise fail.

The refinements in \cite{beccuti2005hybrid} streamline the MLD representation by simplifying logical conditions and replacing nonlinearities with PWA functions. Here, system evolution is governed by affine dynamics over regions defined by inequalities, that is:
\begin{equation}
x^{+} = A_{ti} x + B_{1ti} u + C_{ti}, 
\qquad \text{if } P_{ti} x + Q_{ti} \le 0,\; i = 1,\dots,n
\end{equation}

This reduces mixed-integer complexity retaining the essential event-driven structure. Beyond voltage control, hybrid formulations have been used to coordinate multiple devices across networks \cite{moradzadeh2010hybrid,nguyen2015hybrid} and to combine different hybrid formalisms. For instance, hybrid Petri nets merged with HA \cite{1713273} allow discrete operations and continuous electrical variables to be treated within a unified framework. Recent supervisory strategies for renewable-based systems \cite{4620733,kafetzis2020energy} and distributed-generation environments \cite{https://doi.org/10.1155/2023/5037957} demonstrate the broader applicability of hybrid control, with reported gains such as a 99.4\% reduction in diesel use and improvements of 20\% in voltage stability and 15\% in rotor-angle performance.

\subsubsection{Potential research direction} \label{subsubsec:stability_summary}
 Given the demonstrated values of HA with other modeling frameworks, future research could explore more multi-framework approaches that leverage the strengths of different hybrid models. Refining stability criteria is essential, as many existing models using the notion of hybrid system reachability analysis rely on overly conservative estimates. Current hybrid approaches often focus on grid-connected systems. A further avenue is the development of control strategies for islanded and weak grid conditions, where discrete events and renewable variability interact more strongly and require careful treatment.

\subsection{Anomaly and fault detection and location} \label{subsec:anomaly_detection}
The comprehensiveness achieved in hybrid models of power systems can enable the detection and diagnosis of anomalies that might otherwise go unnoticed when discrete events are triggered.

Early work using HIOA-based formalisms \cite{fourlas2001framework,fourlas2002diagnosability} propose the idea of modeling faults as alternative hybrid modes with distinct guards and transitions. These studies highlight several enduring themes such as the value of representing faulty and healthy behaviors within the same hybrid structure, the importance of analyzing transient modes rather than only steady-state behavior, and the inherent limitations of assuming fail-safe actuators or sensors. Subsequent extension in \cite{Fourlas2005diagnoser} broadens this viewpoint by adding a general hypothesis-testing and removing restrictive fail-safe assumptions, illustrating how hybrid diagnosers can discriminate between multiple fault scenarios even when disturbances produce overlapping behaviors.

Hybrid formulations have also evolved to the addition of control actions directly into the transition logic. Controlled hybrid systems \cite{wang2005hybrid} augment the guard conditions in \eqref{eq: HA} with input-dependent transitions such that:
\begin{equation}
G(q,q') \times V \subseteq X \times U \rightarrow E
\end{equation}
This allows faults identification through both system trajectories and inconsistencies between commanded and observed transitions. 

Uncertainty-aware hybrid methods further expand the achieved fault detection capabilities. Stochastic HA with unscented particle filtering \cite{8535711} demonstrate that probabilistic transitions and hybrid-mode estimation improves robustness to noisy measurements and parameter uncertainty, particularly for systems such as battery management where sensor bias and scaling errors must be isolated. A noticeable development in cyber–physical power systems is the application of hybrid reachability analysis to detect false-data injection attacks, where invariant sets of converter voltages and currents in a hybrid formalism are evaluated against reachable sets to detect not only physical faults but also malicious data manipulation \cite{FDIAattack}.

\subsubsection{Potential research directions} \label{subsubsec:anomaly_summary}

Hybrid fault detection studies demonstrate that complex event sequences and abnormal operating conditions can be analyzed with finer resolution than in traditional frameworks. However, several important challenges remain insufficiently addressed. These include diagnosing multiple or cascading faults, achieving real-time performance under uncertainty, and dealing with cyber induced disruptions. Recent contributions \cite{10441812,7953564,article} address portions of these issues, but substantial methodological gaps persist. Promising directions include fast, scalable hybrid estimators capable of operating under uncertainty. Further, evolving power system topology mandates models that incorporate corrective or mitigating actions rather than detection alone, and a broader treatment of stability phenomena that extend beyond voltages, such as frequency excursions, resonant interactions, and converter driven stability \cite{9286772}.

\subsection{Dynamic performance analysis} \label{subsec:dynamical_analysis}
Hybrid dynamical models have supported a broader understanding of dynamic performance in power systems. Much of this work examines how switching logic, protection timing, and nonlinear generator dynamics shape stability margins and the evolution of cascading outages \cite{Susuki2009,YangEvent}. Event-triggered mechanisms and variable time–stepping have been used to determine optimal timing of protection actions and how transient behaviors should unfold, for an improved analysis of blackout propagation.

Trajectory sensitivity analysis has also been extended to hybrid models. The work in \cite{hiskens2000trajectory} established how sensitivities change when jump events occur, clarifying how small perturbations influence post-disturbance trajectories. These ideas were later extended to second-order sensitivities in \cite{8684267} within the DAIS framework such that curvature effects near switching surfaces and the associated jump conditions are described with improved accuracy. Their result shows that second-order formulation reduces trajectory reconstruction error by 46\%-64\% in all test cases compared to the first-order method.

\subsubsection{Potential research directions} \label{subsubsec:dynamical_summary}
While the literature presents extensive theoretical work on trajectory sensitivity analysis, it often lacks a detailed exploration of the practical challenges involved in applying these methods to large-scale operational grids. A promising future direction is the integration of AI and machine learning, particularly in foundational models tailored for power grids. These models can generalize across system configurations and scale to high-dimensional, multimodal data sources (e.g., synchrophasor, weather, and market data). This will provide computational speed-ups of several orders of magnitude over conventional solvers \cite{osti_2496246}. The sensitivity of event timing should also be further explored, particularly in scenarios where the timing of events has a significant impact on system behavior. For instance, as noted in \cite{8684267}, variations in system parameters can shift the timing of discrete events such as switching actions—an effect that current sensitivity analysis methods may not fully capture.

\subsection{Optimization} \label{subsec:optimization}

Recognizing the need to sustain optimal performance during transitions across different operational states, recent optimization methods integrate discrete events directly into the control formulation. Mixed-integer optimal control problems are  derived from hybrid formalisms \cite{kwatnyship,5721189,thomas2003moving}. These formulations embed mode changes, switching actions, protection triggers, and device limits into a unified structure alongside continuous system dynamics. 

In photovoltaic power systems, \cite{5721189} extends this idea by using the MLD model to construct a mixed-integer MPC controller for maximum power point tracking, closely following the framework of \cite{thomas2003moving}. The resulting problem optimizes both continuous control inputs and binary switching variables subject to the linear mixed-integer constraints defined by the MLD representation.

Hybrid formulations have also been used in operational scheduling and performance enhancement. In a cogeneration plant, the operational scheduling modeled as MLD‐MPC reduces fuel consumption by 5–7\% while meeting electricity and steam demands \cite{Ferrari-Trecate2004}. Neural‐network‐assisted DAIS optimization improves damping and reduces recovery time in stabilizer tuning \cite{baek2008power}. Switched‐system optimization in renewable‐integrated microgrids reduces diesel wear and battery degradation while lowering fuel consumption under variable generation \cite{wu2018}. These results demonstrate that hybrid-model-based optimization can exploit system mode changes to improve performance across a broad range of applications.

\subsubsection{Potential research directions} \label{subsubsec:optimization_summary}
Hybrid model approaches have yet to see widespread application in power systems, not because advanced methods are lacking as new techniques continue to emerge \cite{clark2024optimalcontrolhybridsystems, ALTIN2018128}, but because the domain remains insufficiently explored. Current optimization methods face limitation in handling non-smooth dynamics introduced by discrete events and impulsive HDS. As recently noted in \cite{clark2024optimalcontrolhybridsystems}, the non-injective nature of a reset map where multiple pre-reset states map to a single post-reset state, especially those leading to dimensionality reduction (submersive resets), poses a significant challenge. These effects alsoo create difficulty in identifying a unique optimal trajectory, as discrete events alter both the size and dimensionality of the state space.

\subsection{Verification} \label{subsec:verification_methods}
\begin{figure*}[!t]
\centering
\includegraphics[width=1\linewidth]{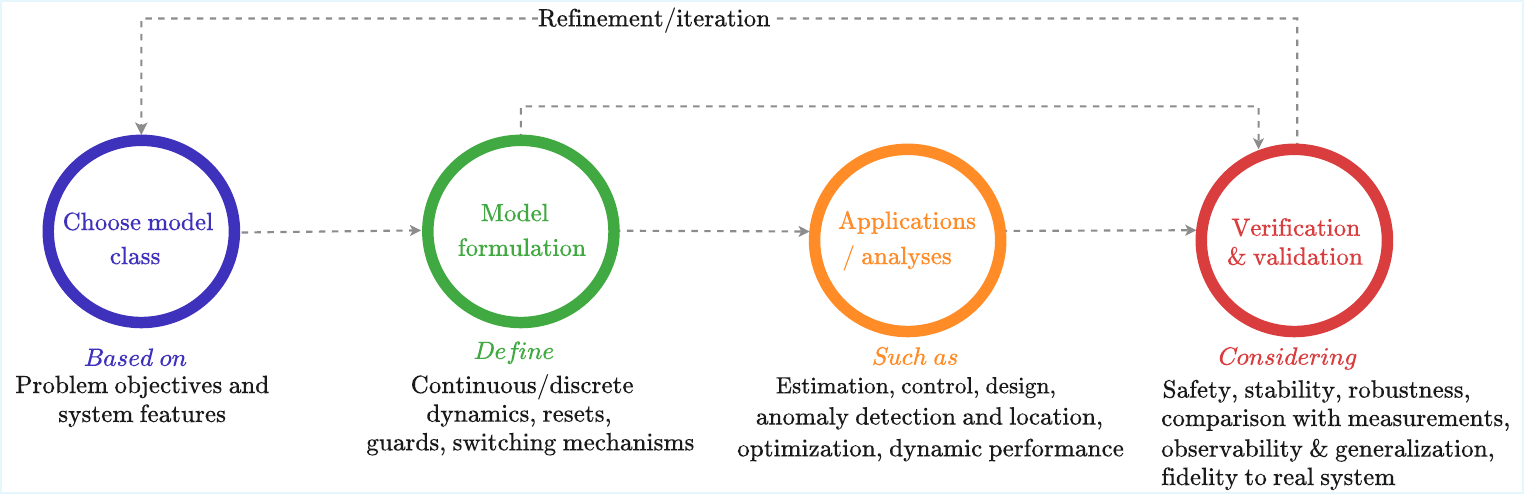}
\caption{Conceptual workflow for HDS modeling in power system applications.}
\label{fig:hdsworkflow}
\end{figure*}

As briefly noted in \autoref{sec:verification}, verification in power systems is challenging because correct operation must be ensured across large state spaces, nonlinear behavior, and diverse operating conditions. These challenges are similar to the long-standing difficulty of verifying hybrid systems in general, which has led to the development of specialized tools for checking safety and correctness. Several of these tools are listed in \autoref{tab:verification-tools}. Within the overall workflow (\autoref{fig:hdsworkflow}), verification and validation form the final step that confirms that hybrid models remain consistent with physical system behavior.

\begin{table}[ht]
\centering
\caption{Verification and simulation tools for hybrid systems (ordered by year of release)}
\label{tab:verification-tools}
\begin{tabular}{p{2.0cm}p{3.9cm}p{1.2cm}}
\toprule
\textbf{Tool} & \textbf{Primary use} & \textbf{Ref.} \\
\midrule
HyTech & Symbolic model checking for linear hybrid automata & \cite{495196} \\
UPPAAL & Verification of timed automata and hybrids & \cite{Larsen1997UppaalIA} \\
HYSDEL & High-level modelling and translation of hybrid systems into various computational networks & \cite{articleHYSDEL} \\
PHAVer & Algorithmic verification for affine systems & \cite{articlePHAver} \\
KeYmaera & Theorem proving for hybrid systems & \cite{platzer2009keymaera} \\
SpaceEx & Verification of affine dynamics & \cite{inproceedingsSpaceEx} \\
Flow* & Reachability analysis for nonlinear hybrid systems & \cite{10.1007/978-3-642-39799-8_18} \\
HyEQ Toolbox & Simulation of hybrid systems in MATLAB/Simulink & \cite{Sanfelice2013} \\
dReach & Bounded model checking for nonlinear hybrid systems & \cite{dReach} \\
HyComp & Verification of hybrid automata with polynomial dynamics & \cite{10.1007/978-3-662-46681-0_4} \\
CORA & Set-based reachability analysis for hybrid systems & \cite{CORA} \\
C2E2 & Simulation-based verification for hybrid systems & \cite{10.1007/978-3-662-46681-0_5} \\
HyST & Model transformation and translation for hybrid systems & \cite{10.1145/2728606.2728630} \\
TuLiP & Controller synthesis for hybrid systems & \cite{dathathri2016tulip} \\
JuliaReach & Reachability analysis of hybrid systems & \cite{10.1145/3302504.3311804} \\
HyPro & Flowpipe construction-based reachability analysis of linear hybrid systems & \cite{SCHUPP2022104945} \\
\bottomrule
\end{tabular}
\end{table}

Hybrid reachability analysis has played an important role in this area. In \cite{Susuki2008}, reachability computations were used to verify a fault release controller in a DMIB system, showing that the controller prevents transient instability under the modeled conditions. Later work such as \cite{inproceedings} extended this idea to power systems with significant uncertainty by using parameter continuation and the grazing phenomenon to assess how system trajectories approach operational limits under variations introduced by renewable generation and cyber enabled control. Stochastic extensions have also been considered in recent studies. Using a stochastic HA, \cite{Shaid2014} with random variations in renewable generation, fault timing, and system responses, and verified that the system avoids unsafe states such as voltage collapse. In \cite{6975219}, hybrid models were combined with temporal logic specifications to automatically generate test cases and evaluate protection system behavior, addressing limitations of simulation centric approaches \cite{Billinton1994, Abate2008}. Similarly, \cite{8245721} proposes differential dynamic logic with hybrid models to verify a flexible load control strategy, enabling reasoning across both information and physical layers.
\subsubsection{Potential research directions} \label{subsubsec:verification_summary}

Hybrid verification methods have shown considerable promise for assessing operational safety in modern power systems. Several challenges remain. Existing methods are typically designed for offline use and cannot support real time assessment, which is increasingly important in converter dominated grids. The strong reliance on synthetic data limits the trust by system operators in these methods and the ability to understand the full complexity of actual system behavior. Finally, translating verified hybrid models into engineering tools remains difficult. Although \cite{Bak2019} provides an automated conversion from hybrid automata to trajectory equivalent Simulink and Stateflow diagrams, this solution is restricted to a single modeling framework. Extending similar automated translation methods to other hybrid formalisms remains an open research problem.

\section{Final remarks} \label{sec:discussion}
The various applications of HDS methods to power systems have shown their diverse potential from multiple perspectives. Since most hybrid formalisms can be applied broadly, we have categorized specific applications by the overarching problems they address. The use of HDS methods has offered considerable advantages in scenarios where purely continuous approaches may fall short. Although a detailed comparative study between continuous and hybrid approaches for specific problems in the literature has yet to be conducted, preliminary evidence suggests that HDS could provide a more robust framework for power system modeling. Admittedly, there are cases where a continuous approach is sufficient to model a system if continuous dynamics predominantly govern the system's behavior, and discrete events have minimal or no impact on the behaviors being investigated, or when the time-driven and event-driven dynamics are not tightly coupled. In such cases, a continuous model can be as conformant and effective as a hybrid one.

\begin{figure*}[ht!]
\centering
\includegraphics[width=0.9\linewidth]{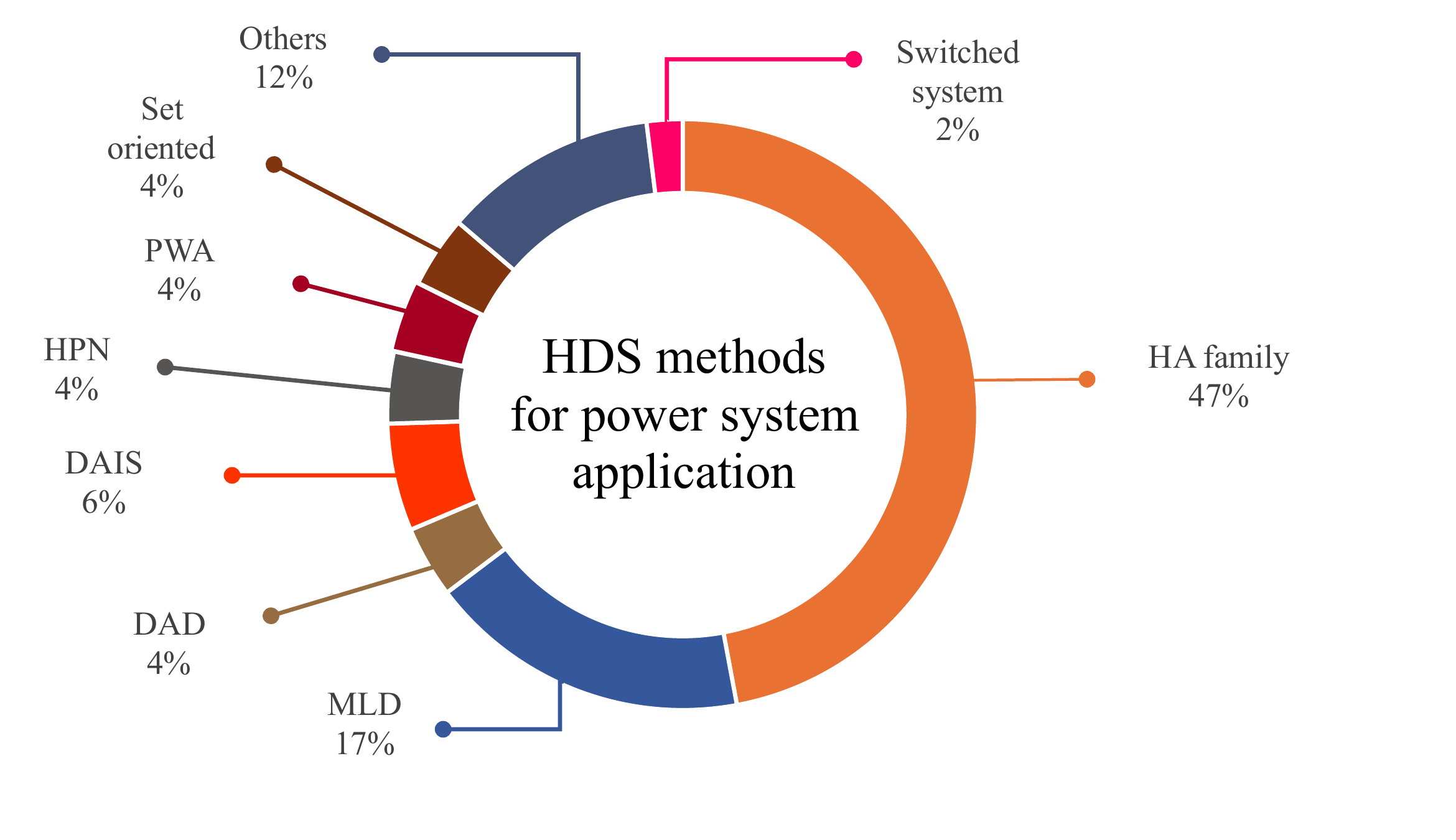} 
\caption{Classification of HDS methods used in power system applications based on the surveyed literature in \autoref{tab:application of HDS research in power system}.}
\label{fig:hdsmethod}
\end{figure*}

As illustrated in \autoref{fig:hdsmethod}, HA-based models account for nearly half of the survey literature, underscoring their prominence in power system applications. This is primarily due to the ease of formulation for small-scale power systems and the formal incorporation of a reset map. Past works typically modify their framework by considering control inputs, disturbances, or integrating probabilistic transition rules. The algorithmic construction of HA and its integration with machine learning methods have also emerged in problems related to control synthesis and verification of hybrid systems. Furthermore, abstracting continuous dynamics in HA, such as DHA, is frequently used to emphasize discrete dynamics, especially when combined with other hybrid frameworks. This approach enhances modeling flexibility tailored to a specific application. However, it is challenging to simulate the HA model with non-linear or non-deterministic guard sets. Conversely, the MLD model is primarily employed in problems relating to optimization, owing to its set of linear inequalities that can easily represent system constraints and combine both continuous and binary variables. Switched system models generally have a straightforward equivalence with PWA systems and can be expressed in MLD form. This flexibility in PWA systems makes it a preferred choice for hybrid model identification in various applications. However, it should be noted that strongly nonlinear behaviors or sharp transitions in power systems may not be well-captured by PWA approximations. 

The robustness of analysis in power systems often relies strongly on the underlying models used. This is why HDS continues to garner attention and sees exciting progress in stability analysis and control. The concept of reachability analysis in hybrid systems is a promising choice for safety verification of power system models, as it can be implemented in both simulation-based and logic-based forms, unlike other methods.

Next, we provide additional directions for further efforts in the broad areas of the HDS approach to modern power system applications.
\begin{itemize}[nosep, leftmargin=.5cm]
\item Comparative studies between continuous and hybrid approaches: There is a need for detailed benchmarking of continuous and HDS approaches in power system modeling. Identifying the specific conditions where hybrid models outperform continuous methods, and vice versa, would provide valuable insights. Such comparisons could also help practitioners select the most suitable modeling framework for different power system analyses.
\item Trajectory Sensitivity under non-differentiable reset maps:
Sensitivity analysis methods in HDS typically assume that reset maps are differentiable, enabling the formulation of jump conditions through analytical derivatives. More recently, these transitions have been systematically characterized using the saltation matrix \cite{saltation}, which captures how infinitesimal perturbations propagate across discrete events. However, renewable-rich power systems often introduce non-differentiable reset behaviors, which pose significant challenges for estimation methods, particularly gradient-based filters. Addressing this gap requires the development of derivative-free methods that can accurately propagate the behavior of non-differentiable reset maps.
\item Enhancing reachability analysis in HDS: There is currently limited application of reachability analysis of HDs to renewable energy models and other applications in power systems. See \autoref{tab:reachable sets in hybrid models of power systems}. Several other methods, such as zonotope-based reachability analysis \cite{7464925,6629354}, ellipsoidal-based reachability analysis \cite{6200393}, symbolic orthogonal projections \cite{BOGOMOLOV2021101093,10.1007/978-3-319-08867-9_27}, and recently with a deep neural network for high-dimensional reachability \cite{deepNNforReachability}, have only been investigated using purely continuous modeling methods for power systems. A detailed comparative study of these methods and those listed in \autoref{tab:reachable sets in hybrid models of power systems} may provide insights into their suitability for specific hybrid model formalisms.
\end{itemize}

\section{Conclusion} \label{sec:conclusion}
This study highlights the growing relevance of HDS methods in power system modeling, particularly in scenarios shaped by high renewable energy penetration and event-driven behaviors. Through a structured review of relevant power system applications ranging from stability analysis and fault diagnosis to controller verification, it becomes evident that HDS approaches offer expressive modeling capabilities that bridge the limitations of traditional continuous-time formulations. A comparative analysis of major hybrid modeling frameworks further clarifies their respective strengths, limitations, and suitability across diverse power system tasks, serving as a reference point for model selection. Despite the advances highlighted in this review, several research gaps remain. These include the need for scalable real-time implementations, improved estimation techniques for systems with non-smooth transitions, and broader adoption of derivative-free sensitivity methods. Addressing these challenges is critical for translating hybrid models into reliable, deployable tools that reflect the operational complexity of renewable-dominated grids. Ultimately, HDS methods provide a powerful framework for developing parsimonious and high-fidelity models that enhance control, optimization, and system awareness in power grids with high renewable penetration. Their ability to represent both continuous dynamics and discrete transitions positions them to significantly impact the design, analysis, and operational reliability of modern power infrastructures which supports the long-term sustainability and resilience of energy systems worldwide.

\section*{Declaration of competing interest}

The authors declare that they have no known competing financial interests or personal relationships that could have appeared to influence the work reported in this paper.

\section*{Data availability}
\noindent
No data were used for the research described in the article.











\end{document}